\documentclass[12pt,english,draftcls, one column]{IEEEtran}

\usepackage[T1]{fontenc}
\usepackage[latin9]{inputenc}
\usepackage{float}
\usepackage{amsmath}
\usepackage{amsthm}
\usepackage{amssymb}
\usepackage{graphicx}
\usepackage{stfloats}
\usepackage{cite}
\usepackage{xcolor}
\usepackage{bbm}
\usepackage{multirow}
\usepackage{multicol}
\usepackage{subfig}
\usepackage{algpseudocode}
\usepackage{ulem}
\usepackage{mathtools}

\newlength{\myeqskip}  
\AtBeginDocument{%
    \setlength\abovedisplayskip{\myeqskip}%
    \setlength\belowdisplayskip{\myeqskip}%
    \setlength\abovedisplayshortskip{\myeqskip-\baselineskip}%
    \setlength\belowdisplayshortskip{\myeqskip}}

\makeatletter

\floatstyle{ruled}
\newfloat{algorithm}{tbp}{loa}
\providecommand{\algorithmname}{Algorithm}
\floatname{algorithm}{\protect\algorithmname}

\theoremstyle{plain}

\theoremstyle{plain}

\IEEEoverridecommandlockouts
\usepackage{cite}
\usepackage{amsfonts}
\usepackage{textcomp}
\usepackage{etoolbox}

\def\BibTeX{{\rm B\kern-.05em{\sc i\kern-.025em b}\kern-.08em
    T\kern-.1667em\lower.7ex\hbox{E}\kern-.125emX}}

\patchcmd{\@maketitle}
  {\addvspace{0.5\baselineskip}\egroup}
  {\addvspace{-1\baselineskip}\egroup}
  {}
  {}

\makeatother

\providecommand{\propositionname}{Proposition}
\providecommand{\theoremname}{Theorem}

\begin{document}

\title{\huge SIM-Enabled Hybrid Digital-Wave Beamforming for Fronthaul-Constrained Cell-Free \\Massive MIMO Systems}

\author{Eunhyuk Park, \textit{Graduate Student Member}, \textit{IEEE}, \\ Seok-Hwan Park, \textit{Senior Member}, \textit{IEEE}, \\ Osvaldo Simeone, \textit{Fellow}, \textit{IEEE}, Marco Di Renzo, \textit{Fellow}, \textit{IEEE}, \\and Shlomo Shamai (Shitz), \textit{Life Fellow}, \textit{IEEE} \thanks{

E. Park and S.-H. Park are with the Division of Electronic Engineering, Jeonbuk
National University, Jeonju, Korea (email: uool$\_$h@jbnu.ac.kr, seokhwan@jbnu.ac.kr).

O. Simeone is with the King's Communications, Learning $\&$ Information Processing (KCLIP) lab within the Centre for Intelligent Information Processing Systems (CIIPS), Department of Engineering, King's College London, London WC2R 2LS, U.K. (email: osvaldo.simeone@kcl.ac.uk).

M. Di Renzo is with Universit\'{e} Paris-Saclay, CNRS, CentraleSup\'{e}lec, Laboratoire des Signaux et Syst\`{e}mes, 3 Rue Joliot-Curie, 91192 Gif-sur-Yvette, France. (marco.di-renzo@universite-paris-saclay.fr), and with King's College London, Centre for Telecommunications Research -- Department of Engineering, WC2R 2LS London, United Kingdom (marco.di\_renzo@kcl.ac.uk).

Shlomo Shamai (Shitz) is with the Department of Electrical and Computer Engineering, Technion Israel Institute of Technology, Haifa 3200003, Israel (e-mail: sshlomo@ee.technion.ac.il).
}
}

\maketitle
\begin{abstract}
As the dense deployment of access points (APs) in cell-free massive multiple-input multiple-output (CF-mMIMO) systems presents significant challenges, per-AP coverage can be expanded using large-scale antenna arrays (LAAs). However, this approach incurs high implementation costs and substantial fronthaul demands due to the need for dedicated RF chains for all antennas.
To address these challenges, we propose a hybrid beamforming framework that integrates wave-domain beamforming via stacked intelligent metasurfaces (SIM) with conventional digital processing. By dynamically manipulating electromagnetic waves, SIM-equipped APs enhance beamforming gains while significantly reducing RF chain requirements. We formulate a joint optimization problem for digital and wave-domain beamforming along with fronthaul compression to maximize the weighted sum-rate for both uplink and downlink transmission under finite-capacity fronthaul constraints. Given the high dimensionality and non-convexity of the problem, we develop alternating optimization-based algorithms that iteratively optimize digital and wave-domain variables. Numerical results demonstrate that the proposed hybrid schemes outperform conventional hybrid schemes, that rely on randomly set wave-domain beamformers or restrict digital beamforming to simple power control. Moreover, the proposed scheme employing sufficiently deep SIMs achieves near fully-digital performance with fewer RF chains in the high signal-to-noise ratios regime.
\end{abstract}



\theoremstyle{plain}
\newtheorem{theorem}{Theorem}
\newtheorem{proposition}{Proposition}
\newtheorem{lemma}{Lemma}
\newtheorem{corollary}{Corollary}
\theoremstyle{definition}
\newtheorem{definition}{Definition}
\theoremstyle{remark}
\newtheorem{remark}{Remark}


\section{Introduction} \label{sec:intro}

\subsection{Background and Motivation} \label{sub:motivation}

Cell-free massive multiple-input multiple-output (CF-mMIMO) systems have emerged as a promising architecture for sixth-generation (6G) wireless networks.
By deploying numerous distributed access points (APs) across a service area, CF-mMIMO systems aim to provide seamless and ubiquitous connectivity to mobile user equipments (UEs) \cite{Ngo:TWC17, Ngo:Proc24}.
These APs are coordinated by a central processor (CP) to enable coherent signal transmission and reception, thereby enhancing interference management.
The performance gains achieved through coherent signal processing among distributed APs have been studied in \cite{Gesbert:JSAC10, Park:TSP13, Zhou:TSP16, Park:TWC16} within the frameworks of network MIMO and cloud radio access networks (C-RAN).

However, the dense deployment of APs in practical scenarios presents significant challenges mainly due to high implementation costs \cite{Jiang:PIMRC24, Gopal:TWC24}. To extend per-AP coverage instead, each AP needs to be equipped with a large-scale antenna array (LAA) \cite{Xu:JSTSP25}. Unfortunately, the system cost increases with the number of radio frequency (RF) chains \cite{Kim:TC19}, making it impractical to assign a dedicated RF chain to every antenna, particularly in LAA-equipped APs.
Additionally, the required fronthaul capacity between APs and CP scales with both the number of antennas and bandwidth \cite{Ngo:SPAWC15}, both of which are expected to increase in 6G, leading to prohibitively high data rate demands on fronthaul links.

\begin{figure}
\centering\includegraphics[width=0.8\linewidth]{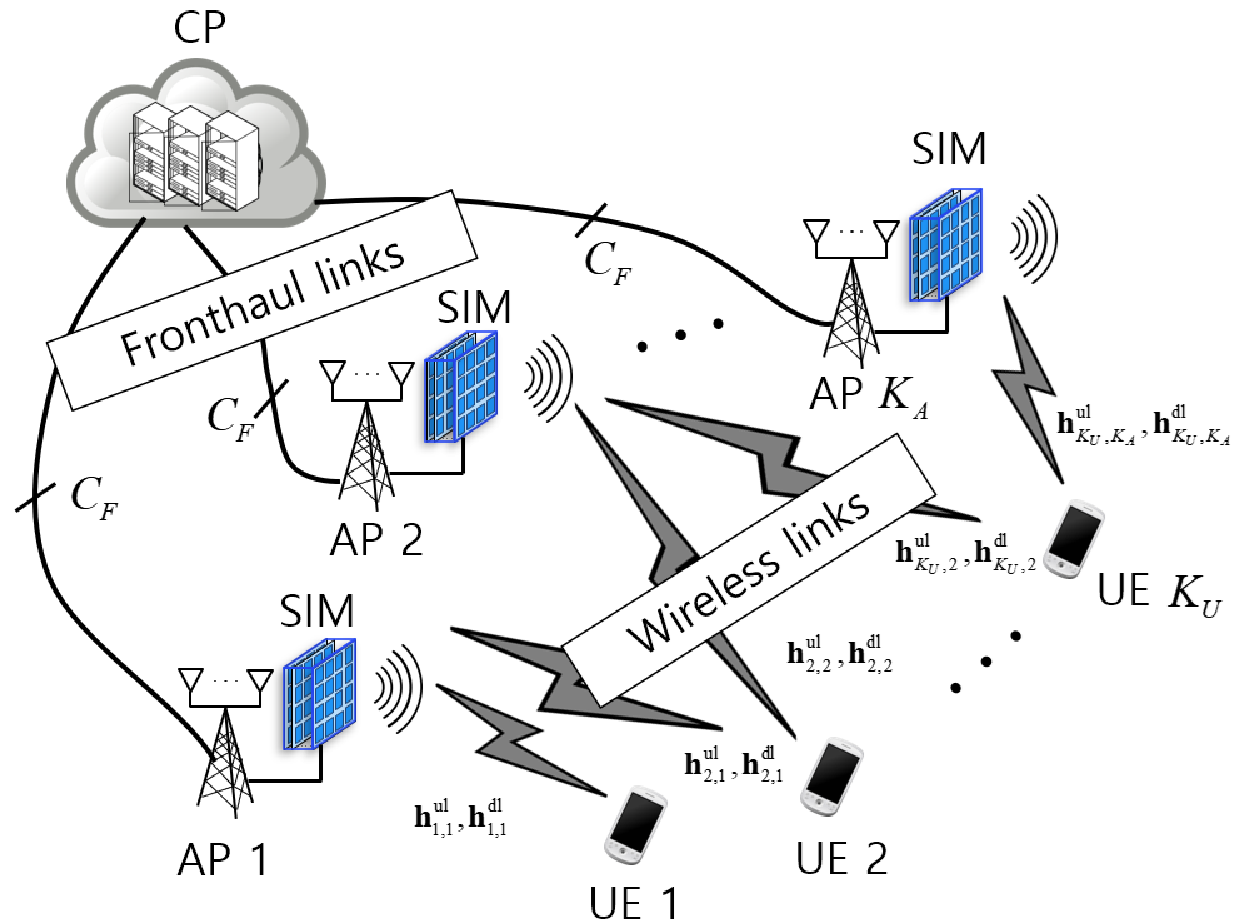}
\caption{\small An SIM-enabled CF-mMIMO system.}
\label{fig:system-model}
\vspace{-3mm}
\end{figure}

To leverage the array gains of LAA-equipped APs in CF-mMIMO systems while utilizing only a limited number of RF chains, as illustrated in Fig. \ref{fig:system-model}, we consider electromagnetic (EM) wave-domain beamforming enabled by stacked intelligent metasurface (SIM).
The SIM architecture consists of multi-layer programmable metasurfaces enclosed in a vacuum container \cite{An:JSAC23, An:ICC23, Papazafeiropoulos:TWC25, Bahingayi:WCL25, Papazafeiropoulos:WCL24, Li:TC24, Papazafeiropoulos:TWC24, Li:TC25, Lin:WCL24, Shi:TWC25-DL, Shi:TWC25-UL, Bahingayi:arXiv25, Hu:TVT25, An:TWC25, Darsena:OJCS25, Shi:arXiv25, Park:PIMRC25}. Each metasurface layer comprises multiple meta-atoms that act as nearly passive elements, dynamically manipulating the phase shift of incoming waves.
By jointly controlling the transmission coefficients of all meta-atoms using a smart controller, such as a field programmable gate array (FPGA) board \cite{An:JSAC23}, APs can perform advanced signal processing directly in the EM wave domain, significantly reducing the reliance on RF chains and the power consumption of analog-to-digital converters which grows with the number of quantization bits and the transmission bandwidth \cite{Lozano:arXiv21}.

To fully exploit SIM-aided CF-mMIMO systems under practical finite-capacity fronthaul constraints, an efficient algorithm is needed for the joint optimization of hybrid digital and wave-domain beamforming, along with fronthaul compression. This optimization is inherently challenging due to the high dimensionality of the solution space. To tackle this challenge, we propose efficient optimization algorithms for both the uplink and downlink of CF-mMIMO systems. By integrating wave-domain processing with conventional digital beamforming and fronthaul compression, our approach enhances system performance while alleviating fronthaul bottlenecks, paving the way for scalable and cost-effective CF-mMIMO deployments. 


\subsection{Related Works} \label{sub:related}

\subsubsection{SIM-Enhanced Wireless Systems} \label{subsub:SIM-Wireless-Systems}

The application of the SIM architecture to single-user MIMO systems was explored in \cite{An:JSAC23,Papazafeiropoulos:TWC24,Bahingayi:WCL25}. In \cite{An:JSAC23}, the optimization of wave-domain beamforming at SIM transceivers was studied with the objective of minimizing the fitting error of the effective channel relative to that generated by a capacity-maximizing singular value decomposition (SVD)-based digital beamformer, while deactivating conventional digital beamforming operation.
Meanwhile, references \cite{Papazafeiropoulos:TWC24} and \cite{Bahingayi:WCL25} focused on directly maximizing the achievable data rate, considering hybrid digital/wave-domain beamforming and pure wave-domain beamforming, respectively.

The impact of SIM on multi-user MIMO systems was investigated in \cite{An:ICC23, Lin:WCL24, Papazafeiropoulos:WCL24, Papazafeiropoulos:TWC25, Li:TC25, An:TWC25, Darsena:OJCS25, Shi:arXiv25, Bahingayi:arXiv25}.
In \cite{Papazafeiropoulos:TWC25}, hybrid digital and wave-domain beamforming was designed for uplink multi-user reception to maximize sum-rate performance.
A low-complexity maximum-ratio combining (MRC) scheme was employed for digital combining, allowing the focus to be placed on optimizing wave-domain beamforming based on the PGA.
In contrast, references \cite{An:ICC23, Lin:WCL24, Papazafeiropoulos:WCL24, An:TWC25, Darsena:OJCS25} studied the downlink of SIM-aided multi-user systems, aiming to maximize sum-rate performance while relying solely on wave-domain beamforming.
In these works, digital-domain processing was limited to power control, which was jointly optimized with wave-domain beamforming using an alternating optimization (AO) approach.
In contrast, references \cite{Li:TC25,Shi:arXiv25,Bahingayi:arXiv25} considered a hybrid digital and wave-domain beamforming to maximize sum-rate or energy efficiency.
While \cite{An:ICC23, An:TWC25, Darsena:OJCS25, Li:TC25, Shi:arXiv25, Bahingayi:arXiv25} assumed the availability of instantaneous channel state information (CSI), \cite{Papazafeiropoulos:TWC25, Lin:WCL24, Papazafeiropoulos:WCL24} relied only on statistical CSI (sCSI).

Recent studies \cite{Li:TC24, Shi:TWC25-UL, Shi:TWC25-DL, Hu:TVT25, Park:PIMRC25} reported that deploying SIMs at APs can enhance the achievable data rates of CF-mMIMO systems for both uplink reception \cite{Li:TC24, Shi:TWC25-UL} and downlink transmission \cite{Shi:TWC25-DL, Hu:TVT25}.
In \cite{Li:TC24}, digital and wave-domain beamforming coefficients at each AP for uplink reception were determined based on local instantaneous CSI, while per-UE central combining vectors at the CP were designed to maximize the signal-to-interference-plus-noise ratio (SINR) using the generalized Rayleigh quotient.
A more practical scenario with only sCSI was considered in \cite{Shi:TWC25-UL}, where all digital and wave-domain beamformers were optimized using sCSI. To this end, a lower bound on the expected per-UE achievable rate was derived, enabling the joint optimization of UEs' transmit powers and APs' wave-domain beamformers under low-complexity MRC local combining at the APs and large-scale fading decoding (LSFD) or equal gain combining decoding (EGCD) schemes at the CP.

For the downlink, \cite{Shi:TWC25-DL} and \cite{Hu:TVT25} focused on wave-domain beamforming design combined with digital-domain power control, excluding digital complex beamforming.
In both studies, each AP antenna was constrained to transmit a single data stream to reduce hardware costs associated with superimposing multiple data streams.
However, fronthaul capacity limitations were not explicitly modeled in these works.
A joint design with digital complex beamforming was studied in \cite{Park:PIMRC25}.

\subsubsection{Fronthaul Compression} \label{subsub:FH-compression}

In CF-mMIMO systems, coherent signal processing across distributed APs is practical only if baseband signals can be reliably exchanged between the APs and CP over fronthaul links with minimal distortion and latency.
However, as both the number of AP antennas and bandwidth increase in 6G systems, fronthaul data rate demands continue to grow, while fronthaul capacity remains limited, making reliable high-speed fronthauling a significant challenge.
Efficient fronthaul compression schemes are therefore essential to transmit key baseband signal information over finite-capacity fronthaul links.

The design of fronthaul compression, alongside digital beamforming, has been explored in several studies, including \cite{Park:TSP13, Zhou:TSP16, Park:TWC16}.
In \cite{Zhou:TSP16}, weighted sum-rate maximization for the uplink was studied under both per-AP independent compression and more advanced Wyner-Ziv compression strategies.
For the downlink, \cite{Park:TSP13} proposed and optimized a multivariate fronthaul compression scheme to maximize weighted sum-rate performance.
While \cite{Park:TSP13} focused on transmitting compressed baseband signals over fronthaul links, \cite{Park:TWC16} introduced a hybrid fronthauling strategy, where each fronthaul link is divided into two sublinks: one for compressed beamformed signals and another for uncoded digital messages.
The optimized hybrid scheme demonstrated significant gains over both pure compression-based and uncoded transmission schemes.

\subsection{Contributions} \label{sub:contribution}


As discussed above, the joint design of digital and wave-domain beamforming, along with fronthaul compression, for both uplink and downlink transmissions remains unaddressed in prior works.
To tackle this challenging problem,
we develop joint optimization algorithms for both uplink and downlink transmission in SIM-aided CF-mMIMO systems.
Given the high-dimensional and non-convex nature of the problems, we develop AO-based algorithms that iteratively optimize digital processing and wave-domain beamforming variables.
Numerical results demonstrate that the hybrid digital-wave schemes optimized using the proposed algorithms outperform conventional hybrid schemes that rely on randomly set wave-domain beamformers or restrict digital beamforming to simple power control.
Moreover, the proposed hybrid schemes employing sufficiently deep SIMs achieve near fully-digital beamforming performance with significantly fewer RF chains in the high signal-to-noise ratio (SNR) regime.

The key contributions are summarized as follows:
\begin{itemize}
    \item We formulate the joint optimization of digital and wave-domain beamforming, along with a fronthaul compression strategy, to maximize the weighted sum-rate for both the uplink and downlink of SIM-aided CF-mMIMO systems under finite-capacity fronthaul constraints.
    \item For the uplink, we develop an AO-based algorithm that alternates between optimizing digital processing variables, involving digital beamforming and fronthaul compression, and wave-domain beamforming variables. To efficiently solve each non-convex subproblem, we employ the matrix Lagrangian duality transform \cite[Thm. 2]{Shen:TN19} and Fenchel's inequality \cite[Lem. 1]{Zhou:TSP16}, leading to convex problems solvable via standard convex solvers.
    \item For the downlink, we adopt a similar AO approach. The
    digital variable subproblem is handled using the same Lagrangian duality and Fenchel's inequality techniques, while the wave-domain subproblem is efficiently solved using a gradient ascent (GA) approach (see, e.g., \cite{Ye:TSP03, Lee:TWC10}), since it is an unconstrained problem.
    \item We present extensive numerical results validating that the proposed hybrid digital-wave schemes achieve significant performance gains over conventional hybrid schemes that rely on randomly fixed wave-domain beamformers or restrict ditigal beamforming to simple power control.
    Furthermore, the proposed schemes employing sufficiently deep SIMs approach the performance of fully-digital beamforming in the high SNR regime.
\end{itemize}


The rest of the paper is organized as follows:
Sec. \ref{sec:System-Model} presents the system model for both uplink and downlink transmission in a CF-mMIMO system, incorporating conventional digital beamforming, SIM-enabled wave-domain beamforming, and fronthaul compression.
The uplink and downlink optimization problems are addressed in Sec. \ref{sec:opt-uplink} and \ref{sec:opt-downlink}, respectively, where AO-based algorithms are developed.
Sec. \ref{sec:numerical} provides extensive numerical results demonstrating the performance gains of the proposed hybrid digital-wave beamforming schemes. Lastly, Sec. \ref{sec:conclusion} concludes the paper.

\textit{Notations:} The complex Gaussian distribution with mean vector $\boldsymbol{\mu}$ and covariance matrix $\boldsymbol{\Sigma}$ is denoted by $\mathcal{CN}(\boldsymbol{\mu}, \boldsymbol{\Sigma})$.
The sets of $M\times N$ complex and real matrices are denoted by $\mathbb{C}^{M\times N}$ and $\mathbb{R}^{M\times N}$, respectively, while $\mathbb{D}^M$ represents the set of $M\times M$ diagonal matrices.
The subset $\mathbb{R}^{M\times N}_+ \subset \mathbb{R}^{M\times N}$ consists of nonnegative real matrices.
The mutual information between random variables $X$ and $Y$ is given by $I(X; Y)$.
The conjugate, transpose, Hermitian transpose and inverse operator are denoted by $(\cdot)^*$, $(\cdot)^T$, $(\cdot)^H$ and $(\cdot)^{-1}$ respectively.
Lastly, $\text{diag}(\cdot)$ returns a diagonal matrix with the input elements as its diagonal, while $\text{blkdiag}(\cdot)$ constructs a block diagonal matrix from the given input matrices.


\section{System Model\label{sec:System-Model}}


As illustrated in Fig. \ref{fig:system-model}, we consider a CF-mMIMO system comprising a CP, $K_A$ APs, and $K_U$ user equipments (UEs).
Each AP is equipped with $N$ antennas, each paired with its own RF chain\footnote{The number of RF chains can be less than $N$ by incorporating analog beamforming (see, e.g., \cite{Sohrabi:JSTSP16}). However, this work focuses on the synergy between digital and wave-domain beamforming, and the joint optimization involving analog beamforming is left for future research.}, while each UE is equipped with a single antenna.
The CP is connected to the APs via error-free digital fronthaul links, each of a finite capacity $C_F$ bps/Hz.
The APs communicate with the UEs over a wireless channel.
While there is no strict constraint on the relationship between $N$ and $K_U$, it is desirable to choose $N$ such that the total number of AP antennas, $K_A N$, is at least $K_U$, i.e., $K_A N \geq K_U$, in order to enable full spatial multiplexing for all $K_U$ UEs.

To ensure ubiquitous connectivity for mobile UEs in CF-mMIMO systems, each AP requires a large number of antennas $N$.
However, this necessitates deploying $N$ dedicated RF chains per AP, leading to high hardware costs.
To address this limitation, we employ SIM-enabled wave-domain beamforming \cite{An:JSAC23, An:ICC23, Papazafeiropoulos:TWC25, Papazafeiropoulos:WCL24, Li:TC24, Papazafeiropoulos:TWC24, Li:TC25, Lin:WCL24, Shi:TWC25-DL, Shi:TWC25-UL, Hu:TVT25, An:TWC25}.
By leveraging well-designed wave-domain beamforming, the system can achieve performance gains while reducing the number of RF chains $N$ required at each AP.

To this end, we assume that each AP is equipped with an SIM positioned between the air interface and its antennas, as shown in Fig. \ref{fig:system-model}.
The SIM at each AP consists of $L$ metasurface layers,
with each layer comprising $M$ meta-atoms.
The wave-domain processing at each metasurface layer will be detailed in the following subsections for both uplink and downlink transmissions.
For notational convenience, we define the following index sets: $\mathcal{K}_A = \{1,2,\ldots,K_A\}$, $\mathcal{K}_U = \{1,2,\ldots,K_U\}$, $\mathcal{N} = \{1,2,\ldots,N\}$, $\mathcal{L} = \{1,2,\ldots,L\}$, and $\mathcal{M} = \{1,2,\ldots,M\}$.

\subsection{Uplink System Model \label{sub:Uplink-model}}

\begin{figure}
\centering\includegraphics[width=0.5\linewidth]{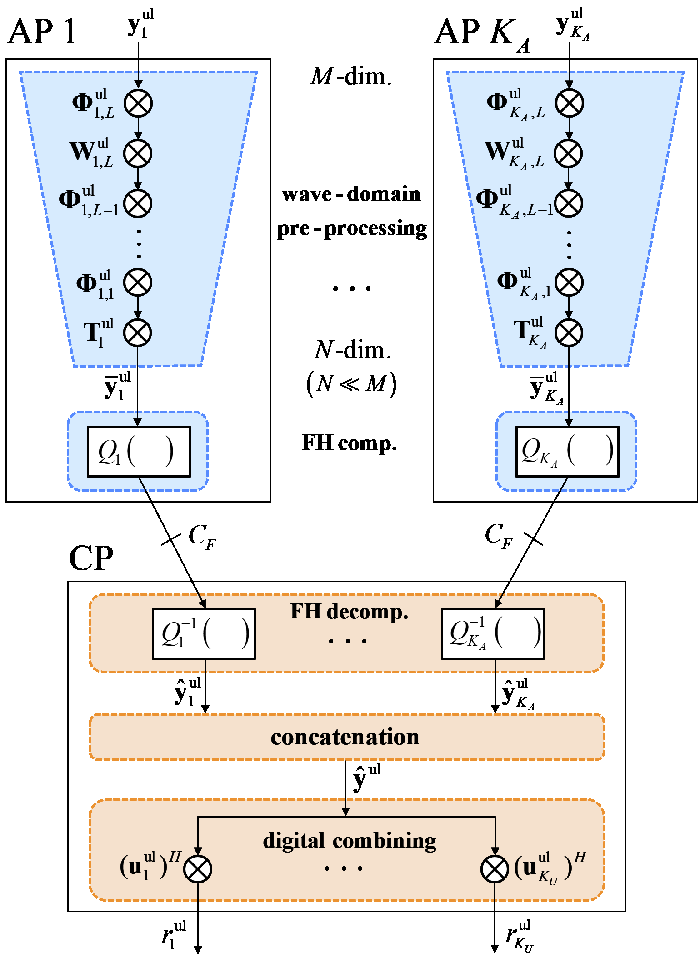}
\caption{\small Uplink signal processing at SIM-enabled APs and CP.} \label{fig:block-diagram-CP-APs-ul}
\vspace{-3mm}
\end{figure}

In uplink data transmission, as shown in Fig. \ref{fig:block-diagram-CP-APs-ul}, each UE $k$ transmits a data signal $s_k^{\text{ul}}\sim\mathcal{CN}(0,p_k^{\text{ul}})$ over a wireless uplink channel, where $p_k^{\text{ul}} \in[0, P_U]$ represents the transmission power with a power budget $P_U$.
The signals received at the APs undergo wave-domain beamforming, fronthaul compression/decompression, and digital combining, as detailed in this subsection.
As the wave-domain beamforming in the uplink occurs prior to digital processing, we refer to it as \textit{wave-domain pre-processing}.
In Fig. \ref{fig:block-diagram-CP-APs-ul}, the fronthaul compression and decompression operators are denoted by $Q_i(\cdot)$ and $Q_i^{-1}(\cdot)$, respectively.

\subsubsection{Uplink Channel and Wave-Domain Pre-Processing} \label{subsub:channel-uplink}

The received signal $\mathbf{y}_i^\text{ul}\in\mathbb{C}^{M\times 1}$ at the input SIM layer of AP $i$ is given by $\mathbf{y}_i^\text{ul} = \sum\nolimits_{k\in\mathcal{K}_U} \mathbf{h}_{k,i}^\text{ul} s_k^{\text{ul}}$, where $\mathbf{h}_{k,i}^\text{ul} \in \mathbb{C}^{M \times 1}$ is the uplink channel vector from UE $k$ to AP $i$.


As illustrated in Fig. \ref{fig:block-diagram-CP-APs-ul}, the signal $\mathbf{y}_i^\text{ul}$ propagates through the SIM deployed at AP $i$, undergoing \textit{wave-domain} pre-processing \cite{An:JSAC23, Papazafeiropoulos:TWC25, An:ICC23}.
The signal $\bar{\mathbf{y}}_i^\text{ul}\in\mathbb{C}^{N \times 1}$ received by $N$ antennas of AP $i$ is a noisy version of the pre-processed signal and is given by
\begingroup
\allowdisplaybreaks
\begin{align}
    \bar{\mathbf{y}}_i^\text{ul} &=    \mathbf{T}_i^{\text{ul}} \boldsymbol{\Phi}_{i, 1}^{\text{ul}} \mathbf{W}_{i, 2}^{\text{ul}} \boldsymbol{\Phi}_{i, 2}^{\text{ul}}  \cdots \boldsymbol{\Phi}_{i, L-1}^{\text{ul}} \mathbf{W}_{i, L}^{\text{ul}} \boldsymbol{\Phi}_{i, L}^{\text{ul}} \, \mathbf{y}_i^\text{ul} + \tilde{\mathbf{z}}_i^\text{ul}, \label{eq:wave-beamforming-ul}
\end{align}
\endgroup
where $\mathbf{T}_i^{\text{ul}}\in\mathbb{C}^{N\times M}$ is the transmission matrix from the output metasurface layer to the $N$ antennas, $\mathbf{W}_{i,l}^{\text{ul}}\in\mathbb{C}^{M\times M}$ represents the transmission matrix between the $l$th and $(l-1)$th metasurface layers, and $\boldsymbol{\Phi}_{i,l}^{\text{ul}} = \text{diag}(\{ e^{j\theta_{i,l,m}^{\text{ul}}} \}_{m\in\mathcal{M}}) \in\mathbb{C}^{M\times M}$ is the transmission coefficient matrix of the $l$th metasurface layer. Here, each $\theta_{i,l,m}^{\text{ul}}\in[0, 2\pi)$ denotes the phase shift applied at the $m$th meta-atom. $\tilde{\mathbf{z}}_i^\text{ul} \sim\mathcal{CN}(0,\sigma_\text{ul}^2\mathbf{I}_N)$ represents the additive noise vector with $\sigma_\text{ul}^2$ denoting the noise variance per antenna.
It is worth noting that the cascade model in (\ref{eq:wave-beamforming-ul}), which comprises inter-layer channels and per-layer phase shifts, is derived under the assumptions of no mutual coupling and a unilateral approximation. For more accurate and generalized SIM architectures, the Z-parameters model proposed in \cite{Abrardo:TC25} can be adopted, as it does not rely on such specific assumptions.

Defining the overall wave-domain pre-processing matrix as $\mathbf{G}_i^{\text{ul}} =  \boldsymbol{\Phi}_{i, 1}^{\text{ul}} \mathbf{W}_{i, 2}^{\text{ul}} \boldsymbol{\Phi}_{i, 2}^{\text{ul}}  \cdots \boldsymbol{\Phi}_{i, L-1}^{\text{ul}} \mathbf{W}_{i, L}^{\text{ul}} \boldsymbol{\Phi}_{i, L}^{\text{ul}} \in \mathbb{C}^{M\times M}$, the received signal in (\ref{eq:wave-beamforming-ul}) simplifies to
\begingroup
\allowdisplaybreaks
\begin{align}
    \bar{\mathbf{y}}_i^\text{ul} = \mathbf{T}_i^{\text{ul}} \mathbf{G}_i^{\text{ul}} \mathbf{y}_i^\text{ul}  + \tilde{\mathbf{z}}_i^\text{ul}
    = \sum\nolimits_{k\in\mathcal{K}_U} \tilde{\mathbf{h}}_{k,i}^\text{ul} s_k^{\text{ul}} + \tilde{\mathbf{z}}_i^\text{ul}, \label{eq:wave-beamforming-process-ul}
\end{align}
\endgroup
where $\tilde{\mathbf{h}}_{k,i}^\text{ul} = \mathbf{T}_i^\text{ul} \mathbf{G}_i^\text{ul} \mathbf{h}_{k,i}^\text{ul}$ is the effective channel between UE $k$ and AP $i$.
It is worth noting that the wave-domain pre-processing in (\ref{eq:wave-beamforming-process-ul}) at the SIM reduces the dimensionality of the received signal vector from $M$ to $N$,
thereby facilitating the fronthaul compression module, described later, in reducing the required compression rate.
Additionally, it is remarked that, as the SIM becomes deeper with a larger $L$, an improved beamforming gain is expected thanks to increased degrees of control in the beamforming design \cite{An:JSAC23, Bahingayi:arXiv25,Hassan:OJCOM24}.

Following Rayleigh-Sommerfeld diffraction theory \cite{Lin:Sci18}, the $(m,m^{\prime})$th element of $\mathbf{W}_{i,l}^{\text{ul}}$ is expressed as
\begin{align}
    \!\!\mathbf{W}_{i,l}^{\text{ul}}(m,m^{\prime}) = \frac{S_i d_{i,\text{Layer}}}{{d_{i,l,m,m'}^{\,2}}} \!\left( \!\frac{1}{2 \pi d_{i,l,m,m'}} \!-\! \frac{j}{\lambda} \right) \!e^{\frac{j 2\pi d_{i,l,m,m'}}{\lambda}}  \label{eq:elements-W-i-l-ul} 
\end{align}
where $S_i$ is the area of each meta-atom, $d_{i,\text{Layer}}$ denotes the spacing between adjacent metasurface layers; $d_{i,l,m,m'}$ represents the transmission distance between the $m'$th meta-atom in the $\left(l-1\right)$th layer and the $m$th meta-atom in the $l$th layer, and $\lambda$ is the wavelength.
Similarly, the $\left(m,n\right)$th element of $\mathbf{T}_i^{\text{ul}}$ can be computed based on the relative positions of the meta-atoms \cite{Lin:Sci18}.
Since the matrices $\mathbf{T}_i^{\text{ul}}$ and $\mathbf{W}_{i,l}^{\text{ul}}$ depend on the fixed geometry of the SIM, they are assumed to be constant and are not subject to optimization in this work.
They could be further optimized by leveraging the emerging technology known as flexible intelligent metasurface \cite{An:TWC25-FIM, Mursia:TC25}.

To highlight the potential advantages of hybrid digital-wave beamforming, jointly designed with fronthaul compression, we focus on the perfect CSI case as in, e.g., \cite{Shi:TWC25-DL}, assuming that the channel vectors between UEs and SIMs can be accurately estimated using hybrid digital-wave domain channel estimators (see, e.g., \cite{Nadeem:WCNC24, Yao:WCL24}).


\subsubsection{Fronthaul Compression} \label{subsub:FH-compression-uplink}

Due to the finite capacity of the fronthaul links, AP $i$ quantizes the wave-domain pre-processed signal $\bar{\mathbf{y}}_i^\text{ul}$ and forwards a compressed bit stream, corresponding to the quantized signal $\hat{\mathbf{y}}_i^\text{ul}$, to the CP.
We model the quantization process using a Gaussian test channel \cite{Park:TSP13, Park:TWC16, Zhou:TSP16}, a special case of standard point-to-point compression model \cite[Ch. 3]{Gamal:Cambridge11}, where  the quantized signal vector $\hat{\mathbf{y}}_i^\text{ul}$ is given by
\begin{align}
    \hat{\mathbf{y}}_i^\text{ul} = \bar{\mathbf{y}}_i^\text{ul} + \mathbf{q}_i^\text{ul}, \label{eq:quantization-model-ul}
\end{align}
with $\mathbf{q}_i^\text{ul} \sim \mathcal{CN}(\mathbf{0}, \boldsymbol{\Omega}_i^\text{ul})$ representing the quantization noise uncorrelated with $\bar{\mathbf{y}}_i^\text{ul}$.
A standard result from source coding theory \cite[Ch. 3]{Gamal:Cambridge11} ensures that $\hat{\mathbf{y}}_i^\text{ul}$ can be reliably decompressed at CP for sufficiently large blocklength, if the condition $I( \bar{\mathbf{y}}_i^\text{ul}; \hat{\mathbf{y}}_i^\text{ul}) \leq C_F$ holds. Under the Gaussian test channel model (\ref{eq:quantization-model-ul}), this condition becomes \cite{Park:TSP13, Park:TWC16, Zhou:TSP16}:
\begingroup
\allowdisplaybreaks
\begin{align}
    & g_i^{\text{ul}}\big(\mathbf{p}^{\text{ul}}, \boldsymbol{\Omega}_i^\text{ul}, \boldsymbol{\theta}^{\text{ul}}\big) \label{eq:FH-capacity-constraint-ul} \\
    &= \log_2\det\left( \sum\nolimits_{k\in\mathcal{K}_U} p_k^\text{ul} \tilde{\mathbf{h}}_{k,i}^\text{ul} (\tilde{\mathbf{h}}_{k,i}^\text{ul})^H + \sigma_\text{ul}^2 \mathbf{I}_N + \boldsymbol{\Omega}_i^\text{ul} \right) \nonumber \\
    &\quad- \log_2\det\left(\boldsymbol{\Omega}_i^\text{ul}\right) \leq C_F, \nonumber
\end{align}
\endgroup
where $\mathbf{p}^\text{ul}=\{ p_k^\text{ul} \}_{k\in\mathcal{K}_U}$ and $\boldsymbol{\theta}^\text{ul} = \{\theta_{i,l,m}^\text{ul}\}_{i\in\mathcal{K}_A, l\in\mathcal{L}, m\in\mathcal{M}}$.

Instead of using the Gaussian test channel-based compressor, which requires a sufficiently large blocklength, we may adopt a uniform scalar quantizer that operates element-wise on each sample of $\bar{\mathbf{y}}_i^{\text{ul}}$. The resulting quantized signal $\hat{\mathbf{y}}_i^{\text{ul}}$ can be approximately modeled using the additive quantization noise model (AQNM) (see, e.g., \cite{Kim:TWC24}).

We note that AP $i$ can apply an additional digital combining operation to the wave-domain pre-processed signal $\bar{\mathbf{y}}_i^{\text{ul}}$ before fronthaul compression, resulting in the quantized signal $\hat{\mathbf{y}}_i^{\text{ul}} = \mathbf{F}_i^{\text{ul}}\bar{\mathbf{y}}_i^{\text{ul}} + \mathbf{q}_i^{\text{ul}}$ with a digital combiner $\mathbf{F}_i^{\text{ul}} \in \mathbb{C}^{N\times N}$.
However, as long as the quantization noise covariance matrix $\boldsymbol{\Omega}_i^{\text{ul}}$ can be optimized, setting the digital combiner to $\mathbf{F}_i^{\text{ul}} = \mathbf{I}_N$ does not cause any loss of optimality \cite{Coso:TWC09, Park:TVT16}.
Therefore, we omit the digital combining process at the APs.

\begin{remark}
    Under the Gaussian test channel model (\ref{eq:quantization-model-ul}), the statistic of the quantization noise $\mathbf{q}_i^{\text{ul}}$ is characterized by its covariance matrix $\boldsymbol{\Omega}_i^{\text{ul}}$.
From an information-theoretic perspective, $\boldsymbol{\Omega}_i^{\text{ul}}$ determines the shape of the quantization regions in the vector quantizer (see, e.g., \cite{Zamir:TIT96}).
For instance, condition (\ref{eq:FH-capacity-constraint-ul}) implies that reducing the distortion (i.e., choosing a smaller $\boldsymbol{\Omega}_i^{\text{ul}}$) increases the mutual information $I(\mathbf{y}_i^{\text{ul}}, \hat{\mathbf{y}}_i^{\text{ul}})$, and hence, demands a higher fronthaul capacity $C_F$.
\end{remark}

\subsubsection{Digital Combining and Achievable Rates} \label{subsub:digital-combining-rate-uplink}

The total quantized signal vector $\hat{\mathbf{y}}^\text{ul} \!\!=\!\! [(\hat{\mathbf{y}}_1^\text{ul})^H \dots (\hat{\mathbf{y}}_{K_A}^\text{ul})^H]^H\!\in\!\mathbb{C}^{N K_A \!\times\! 1}$ received by the CP through the fronthaul links can be expressed as
\begin{align}
    &\hat{\mathbf{y}}^\text{ul} = \sum\nolimits_{k\in\mathcal{K}_U} \tilde{\mathbf{h}}_k^\text{ul} s_k^{\text{ul}} + \bar{\mathbf{z}}^\text{ul} + \bar{\mathbf{q}}^\text{ul}, \label{eq:received-signal-CP}
\end{align}
where $\tilde{\mathbf{h}}_k^\text{ul} \!\!=\!\! [(\tilde{\mathbf{h}}_{k,1}^\text{ul})^{\!H} \!\dots\! (\tilde{\mathbf{h}}_{k,K_A}^\text{ul})^{\!H}]^{\!H}$, $\bar{\mathbf{z}}^\text{ul} \!\!=\!\! [(\tilde{\mathbf{z}}_{1}^\text{ul})^{\!H} \!\dots\! (\tilde{\mathbf{z}}_{K_A}^\text{ul})^{\!H}]^{\!H} $ $\!\sim\! \mathcal{CN}( \mathbf{0}, \sigma_\text{ul}^2\mathbf{I}_{N K_A})$, and $\bar{\mathbf{q}}^\text{ul} \!=\! [\!(\!\bar{\mathbf{q}}_{1}^\text{ul}\!)^{\!H} \!\!\dots\! (\!\bar{\mathbf{q}}_{K_A}^\text{ul}\!)^{\!H}\!]^{\!H} \!\sim\! \mathcal{CN}(\! \mathbf{0}, \bar{\boldsymbol{\Omega}}^\text{\!ul}\!)$ with $\bar{\boldsymbol{\Omega}}^\text{\!ul} \!\!=\!\! \text{blkdiag}(\!\{\!\boldsymbol{\Omega}_i^\text{ul}\!\}_{i\in\mathcal{K}_A}\!)$.

To decode each $s_k^{\text{ul}}$, the CP applies a digital baseband-domain combining to the received quantized signal $\hat{\mathbf{y}}^{\text{ul}}$ using a combining vector $\mathbf{u}_k^{\text{ul}}\in\mathbb{C}^{N K_A \times 1}$.
The CP then decodes $s_k^{\text{ul}}$ based on the combining output $r_k^{\text{ul}} = ( \mathbf{u}_k^{\text{ul}} )^H \hat{\mathbf{y}}^\text{ul}$.
Consequently, the achievable data rate for UE $k$ is given by
\begingroup
\allowdisplaybreaks
\begin{align}
    R_k^{\text{ul}} &=
    f_{k}^{\text{ul}}\big( \mathbf{p}^\text{ul}, \boldsymbol{\Omega}^\text{ul},  \boldsymbol{\theta}^{\text{ul}}, \mathbf{u}^{\text{ul}} \big) \label{eq:achievable-data-rate-ul} \\
    &= \log_2\left( 1 + p_k^\text{ul} \big| \big( \mathbf{u}_k^{\text{ul}} \big)^H\tilde{\mathbf{h}}_k^\text{ul} \big|^2 \,/\, \text{IF}_k^{\text{ul}}\big( \mathbf{p}^\text{ul}, \boldsymbol{\Omega}^\text{ul}, \boldsymbol{\theta}^{\text{ul}}, \mathbf{u}^{\text{ul}} \big) \right), \nonumber
\end{align}
\endgroup
where $\boldsymbol{\Omega}^\text{ul} \!\!=\!\! \{\boldsymbol{\Omega}_i^\text{ul}\}_{i\in\mathcal{K}_A}$ and $\mathbf{u}^{\text{ul}} \!\!=\!\! \{\mathbf{u}_k^{\text{ul}}\}_{k\in\mathcal{K}_U}$.
We have defined the interference-plus-noise power (INP) as $\text{IF}_k^{\text{ul}}(\!\mathbf{p}^\text{ul}\!, \!\boldsymbol{\Omega}^{\text{ul}}\!\!, \!\boldsymbol{\theta}^{\text{ul}}\!\!, \!\mathbf{u}^{\text{ul}}\!) \!= ( \mathbf{u}_k^{\text{ul}} )^H  \big(\sum\nolimits_{k^{\prime}\in\mathcal{K}_U\setminus\{k\}} p_{k^\prime}^{\text{ul}} \tilde{\mathbf{h}}_{k^\prime}^{\text{ul}} (\tilde{\mathbf{h}}_{k^\prime}^{\text{ul}})^H  +  \sigma_\text{ul}^2\mathbf{I}_{N K_A} + \bar{\boldsymbol{\Omega}}^{\text{ul}} \big) \mathbf{u}_k^{\text{ul}}$.


\subsection{Downlink System Model \label{sub:Downlink-model}}

\begin{figure}
\centering\includegraphics[width=0.5\linewidth]{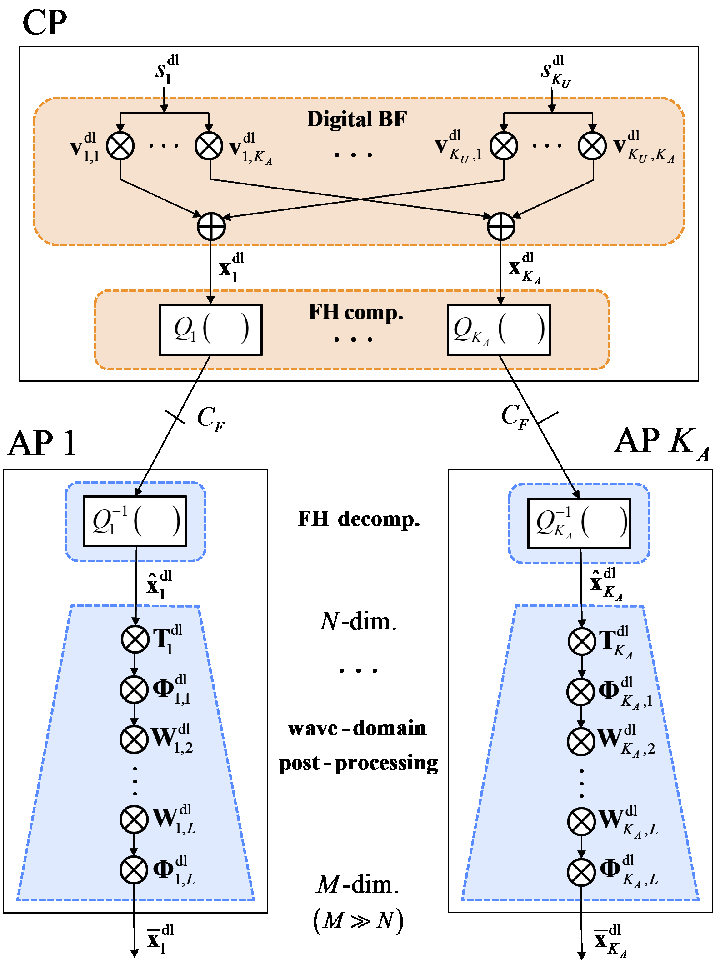}
\caption{\small Downlink signal processing at CP and SIM-enabled APs.} \label{fig:block-diagram-CP-APs-dl}
\vspace{-3mm}
\end{figure}

In downlink data transmission, as illustrated in Fig. \ref{fig:block-diagram-CP-APs-dl}, the data signals $\{s_k^\text{dl}\}_{k\in\mathcal{K}_U}$ intended for the UEs undergo sequential processing through digital beamforming, fronthaul compression/decompression, and wave-domain beamforming, as detailed next.
As the wave-domain beamforming in the downlink is carried out after digital processing, we refer to it as \textit{wave-domain post-processing}.
As in Fig. \ref{fig:block-diagram-CP-APs-ul} for the uplink, $Q_i(\cdot)$ and $Q_i^{-1}(\cdot)$ represent the fronthaul compression and decompression operators, respectively.

\subsubsection{Digital Beamforming} \label{subsub:digital-beamforming-downlink}

At the CP, digital beamforming is applied to the data signals $\{s_k^\text{dl}\}_{k\in\mathcal{K}_U}$, resulting in the precoded signal $\mathbf{x}^\text{dl} = [(\mathbf{x}_1^\text{dl})^H \cdots (\mathbf{x}_{K_A}^\text{dl})^H]^H \in\mathbb{C}^{N K_A \times 1}$ given by
\begin{align}
    \mathbf{x}^\text{dl} = \sum\nolimits_{k\in\mathcal{K}_U} \mathbf{v}_k^\text{dl} s_k^\text{dl}. \label{eq:digital-beamforming}
\end{align}
Here, $\mathbf{v}_k^\text{dl} = [(\mathbf{v}_{k,1}^\text{dl})^H \cdots (\mathbf{v}_{k,K_A}^\text{dl})^H]^H \in \mathbb{C}^{N K_A \times 1}$ denotes the digital beamforming vector for $s_k^\text{dl}$, and the subvectors $\mathbf{x}_i^\text{dl}\in\mathbb{C}^{N\times 1}$ and $\mathbf{v}_{k,i}^\text{dl}\in\mathbb{C}^{N\times 1}$ are the beamformed signal and beamforming vector, respectively, associated with AP $i$.

\subsubsection{Fronthaul Compression} \label{subsub:fronthaul-quantization-downlink}

To enable transmission over the finite-capacity fronthaul links, the CP quantizes each beamformed signal $\mathbf{x}_i^\text{dl}$ and transmits a compressed bit stream representing the quantized signal $\hat{\mathbf{x}}_i^\text{dl}$ to AP $i$.
Similar to the uplink, the quantized signal vector $\hat{\mathbf{x}}_i^\text{dl}$ is modeled as
\begin{align}
    \hat{\mathbf{x}}_i^\text{dl} = \mathbf{x}_i^\text{dl} + \mathbf{q}_i^\text{dl}, \label{eq:quantization-model-dl}
\end{align}
with $\mathbf{q}_i^\text{dl} \sim \mathcal{CN}(\mathbf{0}, \boldsymbol{\Omega}_i^\text{dl})$ representing the quantization noise, uncorrelated with $\mathbf{x}_i^\text{dl}$.
The following constraint needs to be satisfied for a successful decompression of at AP $i$ \cite{Park:TSP13, Park:TWC16, Zhou:TSP16}:
\begin{align}
    &I\left( \mathbf{x}_i^\text{dl} ; \hat{\mathbf{x}}_i^\text{dl}\right) = g_i^\text{dl}\left(\mathbf{v}^\text{dl}, \boldsymbol{\Omega}_i^\text{dl}\right)  \label{eq:FH-capacity-constraint} \\
    &\!=\log_2\det \!\left( \sum\nolimits_{k\in\mathcal{K}_U} \!\!\mathbf{v}_{k,i}^\text{dl}(\mathbf{v}_{k,i}^\text{dl})^H \!\!+\! \boldsymbol{\Omega}_i^\text{dl} \right) \!-\! \log_2\det \! \left(\boldsymbol{\Omega}_i^\text{dl}\right)\! \leq \!C_F, \nonumber
\end{align}
where $\mathbf{v}^\text{dl}=\{ \mathbf{v}_k^\text{dl} \}_{k\in\mathcal{K}_U}$.

We remark that, since the digital beamforming operation (\ref{eq:digital-beamforming}) is applied at the CP, the hardware costs associated with superimposing multiple streams are handled by the CP rather than the APs, and each AP is only required to decompress the quantized version $\hat{\mathbf{x}}_i^{\text{dl}}$ of the digital-beamformed signal.

The transmitted signal vector $\hat{\mathbf{x}}_i^\text{dl}$ from the $N$ antennas of AP $i$ needs to satisfy the following power constraint:
\begin{align}
    \mathbb{E}\left[\|\hat{\mathbf{x}}_i^\text{dl}\|^2\right] = \sum\nolimits_{k\in\mathcal{K}_U} \|\mathbf{v}_{k,i}^\text{dl}\|^2 + \text{tr}\left(\boldsymbol{\Omega}_i^\text{dl}\right) \leq P_A, \label{eq:power-constraint}
\end{align}
with the power budget $P_A$ of AP $i$.

\subsubsection{Wave-Domain Post-Processing} \label{subsub:wave-beamforming-downlink}

As illustrated in Fig. \ref{fig:block-diagram-CP-APs-dl}, the transmitted signal $\hat{\mathbf{x}}_i^\text{dl}$, emitted by the $N$ antennas of AP $i$, passes through the SIM deployed at AP $i$.
The output signal of the wave-domain post-processing $\bar{\mathbf{x}}_i^\text{dl}\in\mathbb{C}^{M \times 1}$ is given by
\begin{align}
    \bar{\mathbf{x}}_i^\text{dl} &= \boldsymbol{\Phi}_{i,L}^{\text{dl}} \mathbf{W}_{i,L}^{\text{dl}} \boldsymbol{\Phi}_{i,L-1}^{\text{dl}} \cdots \boldsymbol{\Phi}_{i,2}^{\text{dl}} \mathbf{W}_{i,2}^{\text{dl}} \boldsymbol{\Phi}_{i,1}^{\text{dl}} \mathbf{T}_i^{\text{dl}} \, \hat{\mathbf{x}}_i^\text{dl}, \label{eq:wave-beamforming}
\end{align}
where $\mathbf{T}_i^{\text{dl}}\in\mathbb{C}^{M\times N}$ denotes the transmission matrix from the $N$ antennas to the input metasurface layer, $\mathbf{W}_{i,l}^{\text{dl}}\in\mathbb{C}^{M\times M}$ represents the transmission matrix between the $(l-1)$th and $l$th metasurface layers, and $\boldsymbol{\Phi}_{i,l}^{\text{dl}} = \text{diag}(\{ e^{j\theta_{i,l,m}^{\text{dl}}} \}_{m\in\mathcal{M}}) \in\mathbb{C}^{M\times M}$ is the phase shift matrix of the $l$th metasurface layer.
The elements of $\mathbf{T}_i^{\text{dl}}$ and $\mathbf{W}_{i,l}^{\text{dl}}$ can be obtained similarly to those in (\ref{eq:elements-W-i-l-ul}) for uplink transmission.
Defining the wave-domain post-processing matrix for downlink transmission as $\mathbf{G}_i^{\text{dl}} = \boldsymbol{\Phi}_{i,L}^{\text{dl}} \mathbf{W}_{i,L}^{\text{dl}} \boldsymbol{\Phi}_{i,L-1}^{\text{dl}} \cdots \boldsymbol{\Phi}_{i,2}^{\text{dl}} \mathbf{W}_{i,2}^{\text{dl}} \boldsymbol{\Phi}_{i,1}^{\text{dl}} \in \mathbb{C}^{M\times M}$, the wave-domain post-processing in (\ref{eq:wave-beamforming}) can be expressed as
\begin{align}
    \bar{\mathbf{x}}_i^\text{dl} = \mathbf{G}_i^{\text{dl}} \mathbf{T}_i^{\text{dl}} \hat{\mathbf{x}}_i^\text{dl}. \label{eq:wave-beamforming-process-dl}
\end{align}
Unlike the uplink wave-domain pre-processing in (\ref{eq:wave-beamforming-process-ul}), which reduces the signal dimension to enable efficient fronthaul compression, the downlink wave-domain post-processing in (\ref{eq:wave-beamforming-process-dl}) expands the dimensionality of the transmitted signal from $N$ to $M$, thereby achieving additional beamforming gain in the downlink channels.

\subsubsection{Downlink Channel and Achievable Rates} \label{subsub:achievable-rate-downlink}

The downlink received signal at UE $k$ is expressed as
\begin{align}
    y_k^\text{dl} = \sum\nolimits_{i\in\mathcal{K}_A} (\mathbf{h}_{k,i}^\text{dl})^H \bar{\mathbf{x}}_i^\text{dl} + z_k^\text{dl}, \label{eq:received-signal-downlink}
\end{align}
where $\mathbf{h}_{k,i}^\text{dl} \in \mathbb{C}^{M\times 1}$ represents the channel vector between the output metasurface layer of AP $i$ and UE $k$, and $z_k^\text{dl}\sim\mathcal{CN}(0, \sigma_\text{dl}^2)$ denotes the additive noise at UE $k$.

For given digital beamforming vectors $\mathbf{v}^\text{dl}$, quantization noise covariance matrices $\boldsymbol{\Omega^\text{dl}}\!\!=\!\! \{\boldsymbol{\Omega}_i^\text{dl}\}_{i\in\mathcal{K}_A}$, and wave-domain post-processing variables $\boldsymbol{\theta}^{\text{dl}} \!\!=\!\! \{\theta_{i,l,m}^{\text{dl}}\}_{i\in\mathcal{K}_A, l\in\mathcal{L}, m\in\mathcal{M}}$,
the SINR can be calculated as
\begin{align}
    &\gamma_k^\text{dl} = \big|(\mathbf{h}_k^\text{dl})^H\bar{\mathbf{G}}^\text{dl} \bar{\mathbf{T}}^\text{dl} \mathbf{v}_k^\text{dl}\big|^2 \, / \, \text{IF}_k^{\text{dl}}\big(\mathbf{v}^\text{dl}, \boldsymbol{\Omega}^\text{dl}, \boldsymbol{\theta}^\text{dl}\big), \label{eq:SINR-dl}
\end{align}
where $\mathbf{h}_k^\text{dl}\!=\![(\!\mathbf{h}_{k,1}^\text{dl}\!)^{\!H} \!\cdots\! (\!\mathbf{h}_{k,K_A}^\text{dl}\!)^{\!H}]^{\!H}$, $\bar{\mathbf{G}}^{\text{dl}} \!=\! \text{blkdiag}(\!\{\mathbf{G}_i^{\text{dl}}\}_{\!i\in\mathcal{K}_A}\!)$, and $\bar{\mathbf{T}}^{\text{dl}} \!=\! \text{blkdiag}(\!\{\mathbf{T}_i^{\text{dl}}\}_{\!i\in\mathcal{K}_{\!A}}\!)$.
The INP at UE $k$ is defined as
\begin{align}
    \text{IF}_k^{\text{dl}} \big(\mathbf{v}^\text{dl}, \boldsymbol{\Omega}^\text{dl}, \boldsymbol{\theta}^{\text{dl}} \big) = \sum\nolimits_{k^{\prime}\in\mathcal{K}_U\setminus\{k\}} \big| (\mathbf{h}_k^\text{dl})^H\bar{\mathbf{G}}^{\text{dl}}\bar{\mathbf{T}}^{\text{dl}}\mathbf{v}_{k^{\prime}}^\text{dl} \big|^2
    + (\mathbf{h}_k^\text{dl})^H \bar{\mathbf{G}}^{\text{dl}} \bar{\mathbf{T}}^{\text{dl}} \bar{\boldsymbol{\Omega}}^{\text{dl}} (\bar{\mathbf{T}}^{\text{dl}})^H (\bar{\mathbf{G}}^{\text{dl}})^H \mathbf{h}_k^\text{dl} + \sigma_\text{dl}^2, \label{eq:dl-INP}
\end{align}
where $\bar{\boldsymbol{\Omega}}^\text{dl} = \text{blkdiag}(\{\boldsymbol{\Omega}_i^\text{dl}\}_{i\in\mathcal{K}_A})$.
Consequently, the achievable data rate of UE $k$ is given by
\begin{align}
    R_k^{\text{dl}} = f_k^{\text{dl}} \big( \mathbf{v}^\text{dl}, \boldsymbol{\Omega}^\text{dl}, \boldsymbol{\theta}^\text{dl} \big) = \log_2\left( 1 + \gamma_k^\text{dl} \right). \label{eq:data-rate-dl}
\end{align}


\section{Uplink Optimization of Hybrid Processing} \label{sec:opt-uplink}

This section discusses the optimization of the uplink hybrid digital-wave processing described in Sec. \ref{sub:Uplink-model}.
We formulate the corresponding optimization problem in Sec. \ref{sub:problem-uplink} and present and evaluate the proposed AO approach to tackle it in Secs. \ref{sub:opt-digital-uplink}--\ref{sub:overall-algorithm-uplink}.



\subsection{Problem Definition} \label{sub:problem-uplink}

We aim at jointly optimizing the power control $\mathbf{p}^\text{ul}$, wave-domain pre-processing $\boldsymbol{\theta}^\text{ul}$, fronthaul compression  $\boldsymbol{\Omega}^\text{ul}$, and digital combining $\mathbf{u}^\text{ul}$ to maximize the weighted sum-rate metric $\sum_{k\in\mathcal{K}_U} \alpha_k^\text{ul} R_k^\text{ul}$.
The problem is formulated as
\begin{subequations} \label{eq:problem-original-ul}
\begin{align}
    \!\!\!\!\underset{\mathbf{p}^\text{ul}, \boldsymbol{\theta}^\text{ul}, \boldsymbol{\Omega}^\text{ul}, \mathbf{u}^\text{ul}} {\mathrm{max.}}\,\,\, & \sum\nolimits_{k\in\mathcal{K}_U} \alpha_k^\text{ul} f_{k}^{\text{ul}} \big( \mathbf{p}^\text{ul}, \boldsymbol{\Omega}^\text{ul}, \boldsymbol{\theta}^\text{ul}, \mathbf{u}^{\text{ul}} \big) \, \label{eq:problem-original-objective-ul} \\
 \mathrm{s.t. }\quad\quad & g_i^{\text{ul}} \big(\mathbf{p}^\text{ul}, \boldsymbol{\Omega}_i^\text{ul}, \boldsymbol{\theta}^\text{ul} \big) \leq C_F, \, \forall i\in\mathcal{K}_A, \label{eq:problem-original-FH-ul} \\
 &  p_k^\text{ul} \in [0, P_U], \, \forall k\in\mathcal{K}_U, \label{eq:problem-original-power-ul} \\
 & \theta_{i,l,m}^\text{ul} \!\in\! [0,2\pi), \, \forall (i,l,m)\in \mathcal{K}_A\!\times\!\mathcal{L}\!\times\!\mathcal{M}. \label{eq:problem-original-wave-ul} 
\end{align}
\end{subequations}
Due to the highly non-convex nature of problem (\ref{eq:problem-original-ul}), we propose an AO algorithm, in which the digital processing variables $\{\mathbf{p}^\text{ul}, \boldsymbol{\Omega}^\text{ul}, \mathbf{u}^\text{ul}\}$ and the wave-domain pre-processing variables $\boldsymbol{\theta}^\text{ul}$ are alternately optimized until convergence.
In the following subsections, we discuss the optimization of $\{\mathbf{p}^\text{ul}, \boldsymbol{\Omega}^\text{ul}, \mathbf{u}^\text{ul}\}$ and $\boldsymbol{\theta}^\text{ul}$ sequentially.


\subsection{Optimization of Digital Processing} \label{sub:opt-digital-uplink}

In this subsection, we discuss the optimization of the digital processing variables $\{\mathbf{p}^\text{ul}, \boldsymbol{\Omega}^\text{ul}, \mathbf{u}^\text{ul}\}$ while keeping the wave-domain pre-processing $\boldsymbol{\theta}^\text{ul}$ fixed.
Even with $\boldsymbol{\theta}^\text{ul}$ given,
the problem (\ref{eq:problem-original-ul}) remains non-convex due to the objective function (\ref{eq:problem-original-objective-ul}) and the fronthaul constraint (\ref{eq:problem-original-FH-ul}). In the following, we describe how to address this non-convexity.

\subsubsection{Handling the Objective Function (\ref{eq:problem-original-objective-ul})}

To handle the non-convexity of the objective function, we employ
the matrix Lagrangian duality transform \cite[Thm. 2]{Shen:TN19}, as presented in the following proposition.

\begin{proposition}\label{prop:matrix-Lagrangian-duality}
Each term $f_{k}^{\text{ul}} \big( \mathbf{p}^\text{ul}, \boldsymbol{\Omega}^\text{ul}, \boldsymbol{\theta}^\text{ul}, \mathbf{u}^{\text{ul}} \big)$ in (\ref{eq:problem-original-objective-ul}) is lower bounded as
\begingroup
\allowdisplaybreaks
\begin{align}
    &f_{k}^{\mathrm{ul}} \big( \mathbf{p}^\mathrm{ul}, \boldsymbol{\Omega}^\mathrm{ul}, \boldsymbol{\theta}^\mathrm{ul}, \mathbf{u}^{\mathrm{ul}} \big) \geq \tilde{f}_{k}^{\mathrm{ul}} \big( \mathbf{p}^\mathrm{ul}, \boldsymbol{\Omega}^\mathrm{ul}, \boldsymbol{\theta}^\mathrm{ul}, \mathbf{u}^{\mathrm{ul}}, \tau_{k}^\mathrm{ul}, \omega_{k}^\mathrm{ul} \big) \nonumber \\
    & = \log_2 \left( 1 \! + \! \tau_{k}^\mathrm{ul} \right) \! - \!  \frac{\tau_{k}^\mathrm{ul}}{\ln2} \! + \! \frac{1+\tau_{k}^\mathrm{ul}}{\ln2} \bigg[ 2\mathrm{Re} \Big\{ \sqrt{p_{k}^\mathrm{ul}} (\tilde{\mathbf{h}}_{k}^\mathrm{ul})^H \mathbf{u}_k^{\mathrm{ul}} \omega_{k}^\mathrm{ul} \Big\} \nonumber \\
    & -\! \big|\omega_{k}^\mathrm{ul}\big|^2 \! \left( p_{k}^\mathrm{ul}\big| \big( \mathbf{u}_k^{\mathrm{ul}} \big)^{\!H} \tilde{\mathbf{h}}_{k}^\mathrm{ul} \big|^2 \! + \! \mathrm{IF}_k^{\mathrm{ul}}\big(\mathbf{p}^\mathrm{ul}, \boldsymbol{\Omega}^{\mathrm{ul}}, \boldsymbol{\theta}^{\mathrm{ul}}, \mathbf{u}^{\mathrm{ul}}\big) \! \right) \! \bigg], \label{eq:convexified-objective-ul}
\end{align}
\endgroup
for any auxiliary variables $\tau_{k}^\mathrm{ul} \in \mathbb{R}_+$ and $\omega_{k}^\mathrm{ul} \in \mathbb{C}$.
The bound in (\ref{eq:convexified-objective-ul}) becomes tight when $\tau_{k}^\mathrm{ul}$ and $\omega_{k}^\mathrm{ul}$ are set as
\begingroup
\allowdisplaybreaks
\begin{subequations} \label{eq:opt-auxiliary-ul}
\begin{align}
    &\tau_{k}^\mathrm{ul} = p_k^{\mathrm{ul}} \big| \big( \mathbf{u}_k^{\mathrm{ul}} \big)^H \tilde{\mathbf{h}}_k^\mathrm{ul} \big|^2 \,/\, \mathrm{IF}_k^{\mathrm{ul}} \big(\tilde{\mathbf{p}}^\mathrm{ul}, \boldsymbol{\Omega}^\mathrm{ul}, \boldsymbol{\theta}^\mathrm{ul}, \mathbf{u}^{\mathrm{ul}} \big), \label{eq:opt-auxiliary-tau-ul} \\
    &\omega_{k}^\mathrm{ul} \!=\!  \sqrt{p_k^{\mathrm{ul}}} \big( \mathbf{u}_k^{\mathrm{ul}} \big)^H \tilde{\mathbf{h}}_k^\mathrm{ul} / \nonumber \\
    &\,\,\,\,\,\,\,\,\,\,\,\,\Big( p_k^{\mathrm{ul}} \big| \big( \mathbf{u}_k^{\mathrm{ul}} \big)^H \tilde{\mathbf{h}}_k^\mathrm{ul} \big|^2 \!+ \mathrm{IF}_k^{\mathrm{ul}} \big(\mathbf{p}^\mathrm{ul}\!\!, \boldsymbol{\Omega}^\mathrm{ul}\!\!, \boldsymbol{\theta}^\mathrm{ul}\!\!, \mathbf{u}^{\mathrm{ul}} \big) \!\Big). \label{eq:opt-auxiliary-omega-ul}
\end{align}
\end{subequations}
\endgroup
\begin{IEEEproof}
Please refer to Appendix \ref{app:proof-matrix-Lagrangian-duality}.
\end{IEEEproof}

\end{proposition}


\subsubsection{Handling the Fronthaul Constraint (\ref{eq:problem-original-FH-ul})}

To mitigate the non-convexity of the fronthaul constraint (\ref{eq:problem-original-FH-ul}), we apply Fenchel's inequality to the $\log_2\det(\cdot)$ function \cite[Lem. 1]{Zhou:TSP16}, leading to the following stricter condition:
\begin{align}
    & \tilde{g}_i^{\text{ul}}\big(\mathbf{p}^\text{ul}, \boldsymbol{\Omega}_i^\text{ul}, \boldsymbol{\theta}^\text{ul},  \boldsymbol{\Xi}_i^\text{ul}\big) = \log_2\det( \mathbf{\Xi}^\text{ul}_i ) \label{eq:convexified-fronthaul-ul} \\
    & \,\,\,+ \frac{1}{\ln 2} \text{tr} \left( (\mathbf{\Xi}_i^\text{ul})^{-1} \! \left( \sum\nolimits_{k\in{\mathcal{K}_U}} p_k^\text{ul} \tilde{\mathbf{h}}_{k,i}^\text{ul} (\tilde{\mathbf{h}}_{k,i}^\text{ul})^H \! + \! \sigma_\text{ul}^2 \mathbf{I}_N \! + \! \boldsymbol{\Omega}_i^\text{ul} \right) \! \right)  \nonumber \\
    & \,\,\, - \frac{N}{\ln 2} - \log_2\det ( \mathbf{\Omega}_i^\text{ul} ) \leq C_F,  \nonumber
\end{align}
where the auxiliary variable $\boldsymbol{\Xi}_i^\text{ul} \succ \mathbf{0}$ is optimally given by
\begin{align}
    \boldsymbol{\Xi}_i^\text{ul} = \sum\nolimits_{k\in{\mathcal{K}_U}} p_k^\text{ul} \tilde{\mathbf{h}}_{k,i}^\text{ul} (\tilde{\mathbf{h}}_{k,i}^\text{ul})^H + \sigma_\text{ul}^2 \mathbf{I}_N + \mathbf{\Omega}_i^\text{ul}, \label{eq:opt-auxiliary-Xi-ul}
\end{align}
which ensures that the constraint (\ref{eq:convexified-fronthaul-ul}) is equivalent to (\ref{eq:problem-original-FH-ul}).

\subsubsection{AO-based Problem Reformulation}

Using the lower bound (\ref{eq:convexified-objective-ul}) and the stricter constraint (\ref{eq:convexified-fronthaul-ul}), we reformulate the optimization problem for the digital processing variables $\{\mathbf{p}^\text{ul}, \boldsymbol{\Omega}^\text{ul}, \mathbf{u}^{\text{ul}}\}$, keeping $\boldsymbol{\theta}^\text{ul}$ fixed, as follows:
\begingroup
\allowdisplaybreaks
\begin{subequations} \label{eq:problem-digital-convexified-ul}
\begin{align}
    \underset{ ^{\mathbf{p}^\text{ul}, \boldsymbol{\Omega}^\text{ul}, \mathbf{u}^{\text{ul}},}_{\boldsymbol{\tau}^\text{ul}, \boldsymbol{\omega}^\text{ul}, \boldsymbol{\Xi}^\text{ul}}} {\mathrm{max.}}\,\,\, & \sum\nolimits_{k\in\mathcal{K}_U} \alpha_k^\text{ul} \tilde{f}_k^{\text{ul}} \left(\mathbf{p}^\text{ul}, \boldsymbol{\Omega}^\text{ul}, \boldsymbol{\theta}^\text{ul}, \mathbf{u}^{\text{ul}}, \tau_k^\text{ul}, \omega_k^\text{ul}\right) \, \label{eq:problem-digital-convexified-objective-ul} \\
    \mathrm{s.t.}\,\,\,\,\,\,\,\, & (\ref{eq:problem-original-power-ul}), (\ref{eq:problem-original-wave-ul}), (\ref{eq:convexified-fronthaul-ul}), \nonumber 
\end{align}
\end{subequations}
\endgroup
where $\boldsymbol{\tau}^\text{ul} \!= \!\{\tau_k^\text{ul}\}_{k\in\mathcal{K}_U}$, $\boldsymbol{\omega}^\text{ul} \!\!=\!\!\{\omega_k^\text{ul}\}_{k\in\mathcal{K}_U}$, and $\boldsymbol{\Xi}^\text{ul} \!\!=\!\! \{\boldsymbol{\Xi}_i^\text{ul}\}_{i\in\mathcal{K}_A}$.

Since the problem (\ref{eq:problem-digital-convexified-ul}) remains non-convex, we partition the optimization variables into three blocks: $\{\mathbf{p}^\text{ul}, \boldsymbol{\Omega}^\text{ul}\}$, $\mathbf{u}^{\text{ul}}$, and $\{\boldsymbol{\tau}^\text{ul},\boldsymbol{\omega}^\text{ul},\boldsymbol{\Xi}^\text{ul}\}$.
When optimizing either $\{\mathbf{p}^\text{ul}, \boldsymbol{\Omega}^\text{ul}\}$ or $\mathbf{u}^{\text{ul}}$ while keeping the remaining variables fixed, the problem becomes convex and can be efficiently solved using optimization tools such as CVX \cite{Grant:CVX20}.
In particular, since the digital combiners $\mathbf{u}^{\text{ul}}$ influence only the objective function through decoupled terms across the UEs, the optimal combiner $\mathbf{u}_k^{\text{ul}}$ for UE $k$ is given as the minimum mean square error (MMSE) combiner:
\begin{align}
    \!\!\mathbf{u}_k^{\text{ul}} = p_k^\text{ul}\bigg( \sum_{k^{\prime}\in\mathcal{K}_U} p_{k^{\prime}}^\text{ul} \tilde{\mathbf{h}}_{k^{\prime}}^\text{ul} (\tilde{\mathbf{h}}_{k^{\prime}}^\text{ul})^H \!+\! \sigma_\text{ul}^2\mathbf{I}_{N K_A} \!+\! \bar{\boldsymbol{\Omega}}^\text{ul}\bigg)^{\!\!-1}\!\tilde{\mathbf{h}}_k^\text{ul}. \label{eq:opt-digital-combiner-ul}
\end{align}
Furthermore, given $\{\mathbf{p}^\text{ul}, \boldsymbol{\Omega}^\text{ul}\}$ and $\mathbf{u}^{\text{ul}}$, the optimal auxiliary variables can be derived in closed form as presented in (\ref{eq:opt-auxiliary-tau-ul}), (\ref{eq:opt-auxiliary-omega-ul}), and (\ref{eq:opt-auxiliary-Xi-ul}).

By leveraging this block-wise structure, we can obtain a sequence of non-decreasing objective values by alternately optimizing
$\{\mathbf{p}^\text{ul}, \boldsymbol{\Omega}^\text{ul}\}$, $\mathbf{u}^{\text{ul}}$, and $\{\boldsymbol{\tau}^\text{ul},\boldsymbol{\omega}^\text{ul},\boldsymbol{\Xi}^\text{ul}\}$. The algorithmic details are presented in Sec. \ref{sub:overall-algorithm-uplink}.


\subsection{Optimization of Wave-Domain Pre-Processing} \label{sub:opt-wave-uplink}

In this subsection, we discuss the optimization of the wave-domain pre-processing $\boldsymbol{\theta}^\text{ul}$ while keeping the digital variables $\{\mathbf{p}^\text{ul}, \boldsymbol{\Omega}^\text{ul}, \mathbf{u}^\text{ul}\}$.
Using (\ref{eq:convexified-objective-ul}) and (\ref{eq:convexified-fronthaul-ul}) similar to Sec. \ref{sub:opt-digital-uplink}, we formulate the optimization problem for the phase shift variables $\boldsymbol{\theta}^\text{ul}$ and the auxiliary variables $\{\boldsymbol{\tau}^\text{ul}, \boldsymbol{\omega}^\text{ul}, \boldsymbol{\Xi}^\text{ul}\}$
given $\{\mathbf{p}^\text{ul}, \boldsymbol{\Omega}^\text{ul}, \mathbf{u}^\text{ul}\}$ as
\begingroup
\allowdisplaybreaks
\begin{subequations} \label{eq:problem-wave-restated-ul}
\begin{align}    \!\!\!\!\underset{\boldsymbol{\theta}^\text{ul}, \boldsymbol{\tau}^\text{ul}, \boldsymbol{\omega}^\text{ul}, \boldsymbol{\Xi}^\text{ul}} {\mathrm{max.}}\,\,\, & \sum\nolimits_{k\in\mathcal{K}_U} \!\alpha_k^\text{ul} \tilde{f}_k^{\text{ul}} \!\left(\mathbf{p}^\text{ul}, \boldsymbol{\Omega}^\text{ul}, \boldsymbol{\theta}^\text{ul}, \mathbf{u}^{\text{ul}}, \tau_k^\text{ul}, \omega_k^\text{ul}\right) \, \label{eq:problem-wave-revisit-objective-ul} \\
    \!\!\mathrm{s.t. }\qquad & \tilde{g}_i^{\text{ul}}\big(\mathbf{p}^\text{ul}, \boldsymbol{\Omega}_i^\text{ul}, \boldsymbol{\theta}^\text{ul},  \boldsymbol{\Xi}_i^\text{ul}\big) \leq C_F, \, \forall i\in\mathcal{K}_A, \label{eq:problem-wave-revisit-FH-ul} \\
    & \theta_{i,l,m}^\text{ul} \in [0,2\pi), \, \forall (i,l,m)\in \mathcal{K}_A\!\times\!\mathcal{L}\!\times\!\mathcal{M}. \label{eq:problem-wave-rivisit-phase-ul}
\end{align}
\end{subequations}
\endgroup
To efficiently solve (\ref{eq:problem-wave-restated-ul}), we employ an AO approach, iteratively updating the wave-domain pre-processing variables $\boldsymbol{\theta}^\text{ul}$ and the auxiliary variables $\{\boldsymbol{\tau}^\text{ul},\boldsymbol{\omega}^\text{ul},\boldsymbol{\Xi}^\text{ul}\}$.
Since the optimal auxiliary variables $\{\boldsymbol{\tau}^\text{ul},\boldsymbol{\omega}^\text{ul},\boldsymbol{\Xi}^\text{ul}\}$ keeping $\boldsymbol{\theta}^\text{ul}$ fixed have closed-form solutions as presented in Sec. \ref{sub:opt-digital-uplink}, we focus on optimizing $\boldsymbol{\theta}^\text{ul}$ while keeping the other fixed.

It is challenging to jointly optimize the phase shift variables $\boldsymbol{\theta}_{i,1}^{\text{ul}}, \boldsymbol{\theta}_{i,2}^{\text{ul}}, \ldots, \boldsymbol{\theta}_{i,L}^{\text{ul}}$ across different layers, given the end-to-end product channel in (\ref{eq:wave-beamforming-ul}).
To address this, we optimize them sequentially in the order $\boldsymbol{\theta}_1^\text{ul}\rightarrow \boldsymbol{\theta}_2^\text{ul} \rightarrow \ldots \rightarrow \boldsymbol{\theta}_L^\text{ul}$, where $\boldsymbol{\theta}_l^\text{ul} = \{\boldsymbol{\theta}_{i,l}^\text{ul}\}_{i\in\mathcal{K}_A}$ collects the phase shift variables of the $l$th layer across all APs.
To further facilitate optimization, we define $\boldsymbol{\Phi}_{i,l}^\text{ul} = \text{diag}(\{ e^{j\theta_{i,l,m}^\text{ul}} \}_{m\in\mathcal{M}})$ and tackle the optimization of each $l$th layer, while keeping the other layers fixed, in terms of $\boldsymbol{\Phi}_l^\text{ul} = \{\boldsymbol{\Phi}_{i,l}^\text{ul}\}_{i\in\mathcal{K}_A}$ instead of $\boldsymbol{\theta}_l^\text{ul}$.
Since $\boldsymbol{\Phi}_l^\text{ul}$ and $\boldsymbol{\Psi}_l^\text{ul}$ have a one-to-one correspondence, we collectively refer to them as the wave-domain pre-processing variables.
The subproblem for the $l$th layer can be stated as
\begingroup
\allowdisplaybreaks
\begin{subequations} \label{eq:problem-wave-wrt-layer-original-ul}
\begin{align}
    \!\!\!\underset{\boldsymbol{\Phi}_l^\text{ul}} {\mathrm{max.}}\,\,\, & \sum\nolimits_{k\in\mathcal{K}_U} \alpha_k^\text{ul} \hat{f}_{k}^{\text{ul}} \left(\boldsymbol{\Phi}_l^\text{ul}, \tau_{k}^\text{ul}, \omega_{k}^\text{ul} \right) \, \label{eq:problem-wave-wrt-layer-original-objective-ul} \\
    \!\!\mathrm{s.t.}\,\,\,\,\, & \hat{g}_i^{\text{ul}}\left(\boldsymbol{\Phi}_{i,l}^\text{ul},  \boldsymbol{\Xi}_i^\text{ul}\right) \leq C_F, \, \forall i\in\mathcal{K}_A, \label{eq:problem-wav-wrt-layer-original-FH-ul} \\
    & \boldsymbol{\Phi}_{i,l}^\text{ul} \in \mathbb{D}^{M}, \, \forall i\in\mathcal{K}_A, \label{eq:problem-wav-wrt-layer-original-Phi-ul} \\
    & \big|\boldsymbol{\Phi}_{i,l}^\text{ul}(m,m)\big| = 1, \, \forall (i,m)\in\mathcal{K}_A\times \mathcal{M}, \label{eq:problem-wav-wrt-layer-original-modulus}
\end{align}
\end{subequations}
\endgroup
where $\boldsymbol{\Phi}_{i,l}^\text{ul}(m,m)$ denotes the $m$th diagonal element of $\boldsymbol{\Phi}_{i,l}^\text{ul}$.
The functions $\hat{f}_{k}^{\text{ul}}$ and $\hat{g}_i^{\text{ul}}$ are defined in (\ref{eq:convexified-wrt-theta-ul}) shown at the top of this page, where the notations $\bar{\boldsymbol{\Phi}}_l^\text{ul} = \text{blkdiag}(\{\boldsymbol{\Phi}_{i,l}^\text{ul}\}_{i\in\mathcal{K}_A})$, $\bar{\mathbf{A}}_{l}^\text{ul} = \text{blkdiag}(\{\mathbf{A}_{i,l}^\text{ul}\}_{i\in\mathcal{K}_A})$ and $\bar{\mathbf{B}}_{l}^\text{ul} = \text{blkdiag}(\{\mathbf{B}_{i,l}^\text{ul}\}_{i\in\mathcal{K}_A})$ are utilized.
Here, the matrices $\mathbf{A}_{i,l}^\text{ul}$ and $\mathbf{B}_{i,l}^\text{ul}$ are given by
\begin{subequations} \label{eq:wave-matrix-ul}
\begin{align}
    &\!\!\!\!\mathbf{A}_{i,l}^\text{ul} \triangleq
    \begin{cases}
        \mathbf{\Phi}_{i,1}^\text{ul} \mathbf{W}_{i,2}^\text{ul} \mathbf{\Phi}_{i,2}^\text{ul} \cdots \mathbf{\Phi}_{i,l-1}^\text{ul} \mathbf{W}_{i,l}^\text{ul}, & \!\!\!\text{if} \,\,l \neq 1, \\
        \mathbf{I}_M, & \!\!\!\text{if} \,\,l=1,
    \end{cases}  \label{eq:wave-matrix-A-ul} \\
    &\!\!\!\!\mathbf{B}_{i,l}^\text{ul} \triangleq
    \begin{cases}
        \mathbf{W}_{i,l+1}^\text{ul} \mathbf{\Phi}_{i,l+1}^\text{ul} \cdots \mathbf{\Phi}_{i,L-1}^\text{ul} \mathbf{W}_{i,L}^\text{ul} \mathbf{\Phi}_{i,L}^\text{ul}, & \!\!\!\text{if} \,\,l \neq L, \\
        \mathbf{I}_M, & \!\!\!\text{if} \,\,l=L.
    \end{cases}  \label{eq:wave-matrix-B-ul}
\end{align}
\end{subequations}

\begin{figure*}[!t]
\begin{subequations} \label{eq:convexified-wrt-theta-ul}
\begin{align}
    &\hat{f}_{k}^{\text{ul}} \left(\boldsymbol{\Phi}_l^\text{ul}, \tau_{k}^\text{ul}, \omega_{k}^\text{ul} \right) = \log_2 \left( 1+\tau_{k}^\text{ul} \right) - \frac{\tau_{k}^\text{ul}}{\ln2} + \frac{1+\tau_{k}^\text{ul}}{\ln2} \bigg[ 2\text{Re} \Big\{ \sqrt{p_{k}^\text{ul}} ( \bar{\mathbf{T}}^\text{ul} \bar{\mathbf{A}}_{l}^\text{ul} \bar{\boldsymbol{\Phi}}_l^\text{ul} \bar{\mathbf{B}}_{l}^\text{ul} \mathbf{h}_{k}^\text{ul}  )^H \mathbf{u}_{k}^{\text{ul}} \omega_{k}^\text{ul} \Big\} \label{eq:convexified-objective-wrt-theta-ul}  \\
    &  -  |\omega_{k}^\text{ul}|^2 \big(\mathbf{u}_{k}^{\text{ul}}\big)^H \left(\bar{\mathbf{T}}^\text{ul} \bar{\mathbf{A}}_{l}^\text{ul} \bar{\boldsymbol{\Phi}}_l^\text{ul} \bar{\mathbf{B}}_{l}^\text{ul} \! \left( \sum\nolimits_{k^{\prime}\in\mathcal{K}_U} p_{k^{\prime}}^\text{ul} \mathbf{h}_{k^{\prime}}^\text{ul} (\mathbf{h}_{k^{\prime}}^\text{ul})^H \! \right)  ( \bar{\mathbf{T}}^\text{ul} \bar{\mathbf{A}}_{l}^\text{ul} \bar{\boldsymbol{\Phi}}_l^\text{ul} \bar{\mathbf{B}}_{l}^\text{ul} )^H \! + \! \sigma_\text{ul}^2 \mathbf{I}_{N K_A} \!+ \!\bar{\boldsymbol{\Omega}}^\text{ul} \right) \mathbf{u}_{k}^{\text{ul}} \bigg], \nonumber \\
    &\hat{g}_i^{\text{ul}}\left(\boldsymbol{\Phi}_{i,l}^\text{ul},  \boldsymbol{\Xi}_i^\text{ul}\right) = \log_2\det( \mathbf{\Xi}_i^\text{ul} ) - \frac{N}{\ln 2} - \log_2\det ( \mathbf{\Omega}_i^\text{ul} ) \label{eq:convexified-fronthaul-wrt-theta-ul} \\
    &  + \frac{1}{\ln 2} \text{tr} \left( (\mathbf{\Xi}_i^\text{ul})^{-1} \left( \mathbf{T}_i^\text{ul} \mathbf{A}_{i,l}^\text{ul} \boldsymbol{\Phi}_{i,l}^\text{ul} \mathbf{B}_{i,l}^\text{ul} \left( \sum\nolimits_{k\in{\mathcal{K}_U}} p_k^\text{ul} \mathbf{h}_{k,i}^\text{ul} (\mathbf{h}_{k,i}^\text{ul})^H \right) ( \mathbf{T}_i^\text{ul} \mathbf{A}_{i,l}^\text{ul} \boldsymbol{\Phi}_{i,l}^\text{ul} \mathbf{B}_{i,l}^\text{ul} )^H + \sigma_{\text{ul}}^2 \mathbf{I}_N + \boldsymbol{\Omega}_i^\text{ul} \right) \right). \nonumber
\end{align}
\end{subequations}
\hrulefill
\vspace{-3mm}
\end{figure*}

The reformulated problem (\ref{eq:problem-wave-wrt-layer-original-ul}) for the $l$th layer remains challenging due to the non-convex unit modulus constraint (\ref{eq:problem-wav-wrt-layer-original-modulus}).
To address this, inspired by the approaches proposed in \cite[Sec. IV-B]{Liu:GC24} and \cite[Sec. III-C]{Zhang:TVT25}, we relax the constraint (\ref{eq:problem-wav-wrt-layer-original-modulus}) to $| \boldsymbol{\Phi}_{i,l}^\text{ul}(m,m)| \leq 1$ and introduce a penalty term into the objective function, leading to the following problem:
\begingroup
\allowdisplaybreaks
\begin{subequations} \label{eq:problem-wave-penalty-original-ul}
\begin{align}
    \!\!\!\underset{ \boldsymbol{\Phi}_l^\text{ul}, \boldsymbol{\Psi}_l^\text{ul}} {\mathrm{max.}}\,\,\, & \!\sum_{k\in\mathcal{K}_U} \!\!\alpha_k^\text{ul} \hat{f}_{k}^{\text{ul}} \left(\boldsymbol{\Phi}_l^\text{ul}, \tau_{k}^\text{ul}, \omega_{k}^\text{ul} \right) \!-\! \xi \!\!\sum_{i\in\mathcal{K}_A} \!\!\big\|\boldsymbol{\Phi}_{i,l}^\text{ul} \!-\! \boldsymbol{\Psi}_{i,l}^\text{ul}\big\|_F^2 \, \label{eq:problem-wave-penalty-original-objective-ul} \\
    \!\!\mathrm{s.t.}\,\,\,\,\,\,\,
    & \hat{g}_i^{\text{ul}}\left(\boldsymbol{\Phi}_{i,l}^\text{ul},  \boldsymbol{\Xi}_i^\text{ul}\right) \leq C_F, \, \forall i\in\mathcal{K}_A, \label{eq:problem-wave-penalty-original-FH-ul} \\
    & \boldsymbol{\Phi}_{i,l}^{\text{ul}}\in\mathbb{D}^M, \,\boldsymbol{\Psi}_{i,l}^{\text{ul}}\in\mathbb{D}^M, \, \forall i\in\mathcal{K}_A,  \label{eq:problem-wave-penalty-original-diagonal-Phi} \\
    &\big| \boldsymbol{\Phi}_{i,l}^\text{ul}(m,m) \big| \leq 1, \, \forall (i,m)\in\mathcal{K}_A\times\mathcal{M}, \label{eq:problem-wave-penalty-original-relaxed} \\
    &\big| \boldsymbol{\Psi}_{i,l}^\text{ul}(m,m) \big| = 1, \, \forall (i,m)\in\mathcal{K}_A\times\mathcal{M}. \label{eq:problem-wave-penalty-original-unit-modulus}
\end{align}
\end{subequations}
\endgroup
Here $\boldsymbol{\Psi}_l^\text{ul} = \{\boldsymbol{\Psi}_{i,l}^\text{ul}\}_{i\in\mathcal{K}_A}$ serves as an auxiliary variable enforcing the unit modulus constraint (\ref{eq:problem-wave-penalty-original-unit-modulus}).
The penalty term in the objective function encourages the wave-domain pre-processing variables $\boldsymbol{\Phi}_l^{\text{ul}}$ to adhere to the constraint (\ref{eq:problem-wav-wrt-layer-original-modulus}) with the penalty coefficient $\xi$ controlling the strength of this enforcement.
The problem (\ref{eq:problem-wave-penalty-original-ul}) can be solved iteratively by alternating updates between the primary variables $\boldsymbol{\Phi}_l^\text{ul}$ and the auxiliary variables $\boldsymbol{\Psi}_l^\text{ul}$.

Assuming $\boldsymbol{\Psi}_l^\text{ul}$ fixed, the optimization over $\boldsymbol{\Phi}_l^\text{ul}$ in (\ref{eq:problem-wave-penalty-original-ul}) becomes convex and can be efficiently solved using standard optimization tools.
Conversely, optimizing $\boldsymbol{\Psi}_l^\text{ul}$ assuming $\boldsymbol{\Phi}_l^\text{ul}$ fixed is a non-convex problem, but it admits a closed-form solution, since it simplifies to
\begin{subequations} \label{eq:problem-wave-penalty-restated-ul-1}
\begin{align}
\!\!\!\underset{{\boldsymbol{\Psi}}_l^\text{ul}} {\mathrm{min.}}\,\,\, & \sum\nolimits_{i\in\mathcal{K}_A}\big\|\boldsymbol{\Phi}_{i,l}^\text{ul} - \boldsymbol{\Psi}_{i,l}^\text{ul}\big\|_F^2 \, \label{eq:problem-wave-penalty-restated-objective-ul-1} \\
    \!\!\mathrm{s.t. }\,\,\,\,
    & \boldsymbol{\Psi}_{i,l}^{\text{ul}} \in \mathbb{D}^{M}, \, \forall i\in\mathcal{K}_A, \label{eq:problem-wave-penalty-restated-diagonal} \\
    &\left|\boldsymbol{\Psi}_{i,l}^\text{ul}(m,m)\right| = 1, \, \forall (i,m)\in\mathcal{K}_A \times \mathcal{M}. \label{eq:problem-wave-penalty-restated-phase-ul-1}
\end{align}
\end{subequations}
Since the objective function in (\ref{eq:problem-wave-penalty-restated-objective-ul-1}) decouples across the diagonal elements as $\sum\nolimits_{i\in\mathcal{K}_A}\|\boldsymbol{\Phi}_{i,l}^\text{ul} - \boldsymbol{\Psi}_{i,l}^\text{ul}\|_F^2 = \sum\nolimits_{i\in\mathcal{K}_A, m\in\mathcal{M}} |\boldsymbol{\Phi}_{i,l}^\text{ul}(m,m) - \boldsymbol{\Psi}_{i,l}^\text{ul}(m,m)|^2$, the solution to (\ref{eq:problem-wave-penalty-restated-ul-1}) is given in closed form as
\begin{align}
    &\boldsymbol{\Psi}_{i,l}^\text{ul} = \text{diag}\left( \big\{ \exp\left( {j\angle \boldsymbol{\Phi}_{i,l}^\text{ul}(m,m)} \right) \big\}_{m\in\mathcal{M}} \right), \label{eq:problem-wave-penalty-closed-form}
\end{align}
for all $i\in\mathcal{K}_A$.
The details of the iterative algorithm are presented in the next subsection.

\begin{remark}
    If each SIM comprises active surface layers, the non-convex unit modulus constraint (\ref{eq:problem-wav-wrt-layer-original-modulus}) is replaced by a convex inequality constraint: $\big|\boldsymbol{\Phi}_{i,l}^\text{ul}(m,m)\big| \leq \phi_{i,l}^{\max}$, where $\phi_{i,l}^{\max} \in (0,1]$ is a fixed bound. Consequently, there is no need to introduce a penalty term or perform a projection step, leading to a more efficient algorithm.
\end{remark}



\subsection{Overall AO Algorithm, Complexity, and Convergence} \label{sub:overall-algorithm-uplink}

\begin{algorithm}
\caption{Proposed AO algorithm for joint optimization of $\{\mathbf{p}^\text{ul},\boldsymbol{\Omega}^\text{ul}, \mathbf{u}^{\text{ul}}\}$ and $\boldsymbol{\theta}^\text{ul}$ for uplink data transmission}\label{algorithm-2}
\begin{algorithmic}[1]
\State \textbf{initialize:}
\State Set $\{\mathbf{p}^\text{ul}, \boldsymbol{\Omega}^\text{ul}\}$ so that the constraints (\ref{eq:problem-original-FH-ul}) and (\ref{eq:problem-original-power-ul}) are satisfied, and initialize $\mathbf{u}^{\text{ul}}$ according to (\ref{eq:opt-digital-combiner-ul}), the phase variables $\boldsymbol{\theta}^\text{ul}$ within $[ 0, 2\pi)$ and the outer iteration count $n^{\text{out}}\leftarrow 1$.
    \Repeat
    \State Set $\xi \leftarrow \xi_0$.
        \Repeat
            \State \textbf{for} $l \in \mathcal{L}$ \textbf{do}
                \State \quad\,\, Update $\{\boldsymbol{\tau}^\text{ul}, \boldsymbol{\omega}^\text{ul}, \boldsymbol{\Xi}^\text{ul}\}$ with (\ref{eq:opt-auxiliary-ul}) and (\ref{eq:opt-auxiliary-Xi-ul}).
                \State \quad\,\, Update $\boldsymbol{\Psi}_l^\text{ul}$ with (\ref{eq:problem-wave-penalty-closed-form}).
                \State \quad\,\, Update $\boldsymbol{\Phi}_l^\text{ul}$ as a solution of the problem (\ref{eq:problem-wave-penalty-original-ul})
                \Statex \qquad\qquad\,\, for fixed $\{\boldsymbol{\Psi}_l^\text{ul}, \boldsymbol{\tau}^\text{ul},\boldsymbol{\omega}^\text{ul},\boldsymbol{\Xi}^\text{ul}\}$.
            \State \textbf{end}
            \State Update $\xi \leftarrow \varrho\xi$.
        \Until {Converged or $n^{\text{wave}} \geq n_{\max}^{\text{wave}}$ \Statex \qquad\quad\,\, (Otherwise, set $n^{\text{wave}} \leftarrow n^{\text{wave}}+1$)}
        \State \textbf{for} $(i,m,l)\in\mathcal{K}_A \times \mathcal{M} \times \mathcal{L}$ \textbf{do}
            \State \quad\,\,$\boldsymbol{\Phi}_{i,l}^\text{ul}(m,m)\leftarrow \exp \left( j\angle \boldsymbol{\Phi}_{i,l}^\text{ul}(m,m) \right)$.
        \State \textbf{end}
        \Repeat
            \State Update $\{\boldsymbol{\tau}^\text{ul},\boldsymbol{\omega}^\text{ul},\boldsymbol{\Xi}^\text{ul}\}$ with (\ref{eq:opt-auxiliary-tau-ul}), (\ref{eq:opt-auxiliary-omega-ul}), and (\ref{eq:opt-auxiliary-Xi-ul}).
            \State Update $\{\mathbf{p}^\text{ul}, \boldsymbol{\Omega}^\text{ul}\}$ as a solution of the problem (\ref{eq:problem-digital-convexified-ul})
            \Statex \qquad\quad for fixed $\{\mathbf{u}^{\text{ul}}, \boldsymbol{\tau}^\text{ul},\boldsymbol{\omega}^\text{ul},\boldsymbol{\Xi}^\text{ul}\}$.
            \State Update $\mathbf{u}^{\text{ul}}$ with (\ref{eq:opt-digital-combiner-ul}).
        \Until {Converged or $n^{\text{digital}} \geq n_{\max}^{\text{digital}}$
        \Statex \qquad\quad\,\, (Otherwise, set $n^{\text{digital}} \leftarrow n^{\text{digital}}+1$)}
    \Until {Converged or $n^{\text{out}} \geq n_{\max}^{\text{out}}$ \Statex \qquad (Otherwise, set $n^{\text{out}} \leftarrow n^{\text{out}}+1$)}
\end{algorithmic}
\end{algorithm}


\subsubsection{Overall AO Algorithm}

The overall AO algorithm is summarized in Algorithm 1, where the optimization of digital processing variables $\{\mathbf{p}^\text{ul}, \boldsymbol{\Omega}^\text{ul}, \mathbf{u}^{\text{ul}}\}$ and wave-domain pre-processing variables $\boldsymbol{\theta}^\text{ul}$ is carried out alternately in Steps 16--20 and 5--12, respectively.
To ensure stable convergence for wave-domain pre-processing optimization, the penalty coefficient $\xi>0$ is gradually increased in each inner iteration according to the rule $\xi\leftarrow\varrho\xi$ with $\varrho > 1$ \cite{Hua:TWC24}.
Additionally, once the wave-domain pre-processing optimization is complete, the obtained $\{\boldsymbol{\Phi}_{l}^{\text{ul}}\}_{l\in\mathcal{L}}$ is projected onto the feasible set to enforce the unit modulus constraint (\ref{eq:problem-wav-wrt-layer-original-modulus}) in Steps 13--15.

\subsubsection{Complexity}

The complexity $C_{\text{total}}^{\text{ul}}$ of Algorithm 1 is given by $C_{\text{total}}^{\text{ul}} = I_{\text{out}}^{\text{ul}} ( C_{\text{digital}}^{\text{ul}} + C_{\text{wave}}^{\text{ul}} )$, where
$C_{\text{digital}}^{\text{ul}}$ and $C_{\text{wave}}^{\text{ul}}$ represent the complexities of digital and wave-domain pre-processing optimization steps, respectively, and $I_{\text{out}}^{\text{ul}}$ denotes the number of outer iterations required for convergence.
The complexity $C_{\text{digital}}^{\text{ul}}$ associated with optimizing the digital processing variables $\{{\mathbf{p}^{\text{ul}}}, \boldsymbol{\Omega}^{\text{ul}}, \mathbf{u}^{\text{ul}}\}$
is given by the product of the number of inner iterations and the complexity of each iteration.
The per-iteration complexity is dominated by the complexity of solving the convex problem (\ref{eq:problem-digital-convexified-ul}) for fixed $\{\mathbf{u}^{\text{ul}}, \boldsymbol{\tau}^{\text{ul}},\boldsymbol{\omega}^{\text{ul}},\boldsymbol{\Xi}^{\text{ul}}\}$.
This complexity is upper bounded by
$\mathcal{O}( n_V^{\text{ul, digital}} ((n_V^{\text{ul, digital}})^3 + n_O^{\text{ul, digital}}) )$ \cite[p. 4]{BTal:LN19},
where $n_V^{\text{ul, digital}} = \mathcal{O}(K_U + N^2 K_A)$ and $n_O^{\text{ul, digital}} = \mathcal{O}(K_A N^2 ( K_A K_U^2 + N ))$ denote the respective numbers of optimization variables and arithmetic operations needed for evaluating the objective and constraint functions, respectively.

The complexity $C_{\text{wave}}^{\text{ul}}$ for optimizing the wave-domain pre-processing variables $\boldsymbol{\theta}^\text{ul}$ is given by the number of inner iterations multiplied by the per-iteration complexity which is dominated by the complexity of solving the convex problem (\ref{eq:problem-wave-penalty-original-ul}) for fixed $\{\boldsymbol{\Psi}^{\text{ul}}, \boldsymbol{\tau}^{\text{ul}},\boldsymbol{\omega}^{\text{ul}},\boldsymbol{\Xi}^{\text{ul}}\}$, where $\boldsymbol{\Psi}^{\text{ul}} = \{\boldsymbol{\Psi}_l^{\text{ul}}\}_{l\in\mathcal{L}}$, with an upper bound given by $\mathcal{O}( n_V^{\text{ul, wave}} ((n_V^{\text{ul, wave}})^3 + n_O^{\text{ul, wave}}) )$, where the numbers of optimization variables and arithmetic operations scale as $n_V^{\text{ul, wave}} = \mathcal{O}( M^2K_A )$ and $n_O^{\text{ul, wave}} = \mathcal{O}( K_AK_U ( NM^2K_A^2 + N^2K_AK_U + LM^3 ) )$, respectively.

\subsubsection{Convergence}

Both the subalgorithms for optimizing the digital processing variables $\{\mathbf{p}^\text{ul}, \boldsymbol{\Omega}^\text{ul}, \mathbf{u}^{\text{ul}}\}$ (Steps 16--20) and wave-domain pre-processing variables $\boldsymbol{\theta}^\text{ul}$ (Steps 5--12) adopt the FP approach, whose convergence to stationary points was established in \cite{Shen:TN19}. Specifically, they ensure a monotonic increase in the objective function with respect to the number of inner iterations.
However, due to the projection operation in Steps 13--15, which is applied after optimizing $\boldsymbol{\theta}^\text{ul}$, a monotonic increase in the objective function across the outer iterations is not mathematically guaranteed.
Nevertheless, as will be illustrated numerically in Sec. \ref{sub:convegence}, Algorithm 1 achieves a monotonically increasing objective function across the outer iterations and converges rapidly in practice.


\section{Downlink Optimization of Hybrid Processing} \label{sec:opt-downlink}


In this section, we address the optimization of the downlink hybrid digital-wave processing described in Sec. \ref{sub:Downlink-model}.
The optimization problem is formulated in Sec. \ref{sub:problem-downlink} and solved using an AO algorithm, which is detailed in Secs. \ref{sub:opt-digital-downlink}--\ref{sub:overall-algorithm-downlink}.


\subsection{Problem Definition} \label{sub:problem-downlink}

Similar to the uplink, we aim at maximizing the weighted sum-rate $\sum_{k\in\mathcal{K}_U} \alpha_k^\text{dl} R_k^{\text{dl}}$ by optimizing the digital beamforming $\mathbf{v}^\text{dl}$, the fronthaul compression $\boldsymbol{\Omega}^\text{dl}$, and the wave-domain post-processing $\boldsymbol{\theta}^\text{dl}$.
The problem is formulated as
\begin{subequations} \label{eq:problem-original}
\begin{align}
    \underset{\mathbf{v}^\text{dl}, \boldsymbol{\Omega}^\text{dl}, \boldsymbol{\theta}^\text{dl}} {\mathrm{max.}}\,\,\, & \sum\nolimits_{k\in\mathcal{K}_U} \alpha_k^\text{dl} f_k^{\text{dl}}\big(\mathbf{v}^\text{dl}, \boldsymbol{\Omega}^\text{dl}, \boldsymbol{\theta}^\text{dl}\big) \, \label{eq:problem-original-objective} \\
 \mathrm{s.t. }\,\,\,\,\,\,\,\, & g_i^\text{dl}\left(\mathbf{v}^\text{dl}, \boldsymbol{\Omega}_i^\text{dl}\right) \leq C_F, \, \forall i\in\mathcal{K}_A, \label{eq:problem-original-FH} \\
 &  \sum\nolimits_{k\in\mathcal{K}_U} \! \|\mathbf{v}_{k,i}^\text{dl}\|^2 \! + \! \text{tr}\left(\boldsymbol{\Omega}_i^\text{dl}\right) \leq P_A, \, \forall i\in\mathcal{K}_A, \label{eq:problem-original-power} \\
 & \theta_{i,l,m}^\text{dl} \in [0,2\pi), \, \forall (i,l,m)\in \mathcal{K}_A\times\mathcal{L}\times\mathcal{M}. \label{eq:problem-original-wave-range} 
\end{align}
\end{subequations}
Since the problem (\ref{eq:problem-original}) is non-convex, we propose an AO algorithm that alternately optimizes the digital processing variables $\{\mathbf{v}^\text{dl}, \boldsymbol{\Omega}^\text{dl}\}$ and the wave-domain post-processing variables $\boldsymbol{\theta}^\text{dl}$.
The optimization of each set of variables, given the other, is discussed in the following subsections.


\subsection{Optimization of Digital Processing} \label{sub:opt-digital-downlink}


In this subsection, we tackle the optimization of the digital processing variables $\{\mathbf{v}^\text{dl}, \boldsymbol{\Omega}^\text{dl}\}$ keeping the wave-domain post-processing $\boldsymbol{\theta}^\text{dl}$ fixed.
Problem (\ref{eq:problem-original}) remains non-convex even keeping $\boldsymbol{\theta}^\text{dl}$ fixed, due to the objective function (\ref{eq:problem-original-objective}) and the fronthaul constraint (\ref{eq:problem-original-FH}).
Similar to the uplink approach in Sec. \ref{sec:opt-uplink}, we address this non-convexity using the matrix Lagrangian duality transform \cite[Thm. 2]{Shen:TN19} and Fenchel's inequality \cite[Lem. 1]{Zhou:TSP16}, as detailed next.

\subsubsection{Handling the Objective Function (\ref{eq:problem-original-objective})} \label{subsub:handling-objective-downlink}

To address the non-convexity of the objective function,
we derive a lower bound on each term $f_k^{\text{dl}}(\mathbf{v}^\text{dl}, \boldsymbol{\Omega}^\text{dl}, \boldsymbol{\theta}^\text{dl})$ using the matrix Lagrangian duality transform \cite[Thm. 2]{Shen:TN19}:
\begingroup
\allowdisplaybreaks
\begin{align}
    & f_k^{\text{dl}}\big( \mathbf{v}^\text{dl}, \boldsymbol{\Omega}^\text{dl}, \boldsymbol{\theta}^\text{dl} \big) \geq \tilde{f}_k^{\text{dl}}\big(\mathbf{v}^\text{dl}, \boldsymbol{\Omega}^\text{dl}, \boldsymbol{\theta}^\text{dl}, \tau_k^\text{dl}, \omega_k^\text{dl}\big) \nonumber \\
    & = \log_2 \left( 1+\,\tau_k^\text{dl} \right) - \frac{\tau_k^\text{dl}}{\ln2} + \frac{1+\tau_k^\text{dl}}{\ln2} \bigg[ 2\text{Re} \Big\{ (\mathbf{v}_k^\text{dl})^H \tilde{\mathbf{h}}_k^\text{dl} \omega_k^\text{dl} \Big\} \nonumber \\
    & \quad - | \omega_k^\text{dl} |^2 \left( \big| (\tilde{\mathbf{h}}_k^\text{dl})^H \mathbf{v}_{k}^\text{dl} \big|^2 + \text{IF}_k^\text{dl}\big(\mathbf{v}^\text{dl}, \boldsymbol{\Omega}^\text{dl}, \boldsymbol{\theta}^\text{dl}\big) \right) \bigg], \label{eq:convexified-objective-dl}
\end{align}
\endgroup
where $\tilde{\mathbf{h}}_k^\text{dl} = (\bar{\mathbf{T}}^\text{dl})^H (\bar{\mathbf{G}}^\text{dl})^H \mathbf{h}_k^\text{dl}$ represents the effective channel toward UE $k$ given the wave-domain post-processing variables.
The lower bound in (\ref{eq:convexified-objective-dl}) becomes equal to $f_k^{\text{dl}}(\mathbf{v}^\text{dl}, \boldsymbol{\Omega}^\text{dl}, \boldsymbol{\theta}^\text{dl})$ when the auxiliary variables $\tau_k^\text{dl} \in \mathbb{R}_+$ and $\omega_k^\text{dl} \in \mathbb{C}$ are set to
\begingroup
\allowdisplaybreaks
\begin{align}
    &\tau_k^\text{dl} = \big|
    (\tilde{\mathbf{h}}_k^\text{dl})^H \mathbf{v}_k^\text{dl} \big|^2 \, / \, \text{IF}_k^\text{dl}\big(\mathbf{v}^\text{dl}, \boldsymbol{\Omega}^\text{dl}, \boldsymbol{\theta}^\text{dl}\big), \label{eq:opt-auxiliary-tau-dl} \\
    &\omega_k^\text{dl} = (\tilde{\mathbf{h}}_k^\text{dl})^H \mathbf{v}_k^\text{dl} / \Big(\big|(\tilde{\mathbf{h}}_k^\text{dl})^H  \mathbf{v}_k^\text{dl} \big|^2 + \text{IF}_k^\text{dl}\big(\mathbf{v}^\text{dl}, \boldsymbol{\Omega}^\text{dl}, \boldsymbol{\theta}^\text{dl}\big) \Big). \label{eq:opt-auxiliary-omega-dl}
\end{align}
\endgroup

\subsubsection{Handling the Fronthaul Constraint (\ref{eq:problem-original-FH})} \label{subsub:handling-fronthaul-constraint-downlink}

Applying Fenchel's inequality \cite[Lem. 1]{Zhou:TSP16}, we derive a stricter constraint that ensures the satisfaction of the fronthaul constraint (\ref{eq:problem-original-FH}):
\begin{align}
    & \tilde{g}_i^{\text{dl}} \left( \mathbf{v}^\text{dl}, \boldsymbol{\Omega}_i^\text{dl}, \boldsymbol{\Xi}_i^{\text{dl}} \right) \!=\! \log_2\!\det( \mathbf{\Xi}_i^{\text{dl}} ) \!-\! \frac{N}{\ln 2} \!-\! \log_2\!\det ( \mathbf{\Omega}_i^\text{dl} ) \label{eq:convexified-fronthaul} \\
    & \,\,\,+ \frac{1}{\ln 2} \text{tr} \left( (\mathbf{\Xi}_i^{\text{dl}})^{-1} \left( \sum\nolimits_{k\in{\mathcal{K}_U}} \mathbf{v}_{k,i}^\text{dl} (\mathbf{v}_{k,i}^\text{dl})^H + \mathbf{\Omega}_i^\text{dl} \right) \right) \leq C_F,  \nonumber
\end{align}
with an auxiliary variable $\boldsymbol{\Xi}_i^{\text{dl}} \succ \mathbf{0}$.
This reformulated constraint (\ref{eq:convexified-fronthaul}) becomes equivalent to (\ref{eq:problem-original-FH}), when
\begin{align}
    \boldsymbol{\Xi}_i^{\text{dl}} = \sum\nolimits_{k\in{\mathcal{K}_U}} \mathbf{v}_{k,i}^\text{dl} (\mathbf{v}_{k,i}^\text{dl})^H + \mathbf{\Omega}_i^\text{dl}. \label{eq:opt-auxiliary-Xi-dl}
\end{align}

\subsubsection{AO-based Problem Reformulation} \label{subsub:reformulating-downlink}

Using the lower bound in (\ref{eq:convexified-objective-dl}) and the stricter constraint in (\ref{eq:convexified-fronthaul}), we reformulate the optimization of the digital processing variables $\{\!\mathbf{v}^\text{dl}\!\!, \boldsymbol{\Omega}^\text{dl}\!\}$ as
\begingroup
\allowdisplaybreaks
\begin{subequations} \label{eq:problem-digital-restated}
\begin{align}
    \!\!\!\underset{ ^{\mathbf{v}^\text{dl}, \boldsymbol{\Omega}^\text{dl},\boldsymbol{\tau}^{\text{dl}}, \boldsymbol{\omega}^{\text{dl}}, \boldsymbol{\Xi}^{\text{dl}}}} {\mathrm{max.}}\,\,\, & \sum\nolimits_{k\in\mathcal{K}_U} \alpha_k^\text{dl} \tilde{f}_k^{\text{dl}}\left(\mathbf{v}^\text{dl}, \boldsymbol{\Omega}^\text{dl}, \boldsymbol{\theta}^{\text{dl}}, \tau_k^{\text{dl}}, \omega_k^{\text{dl}}\right) \, \label{eq:problem-digital-convexified-objective} \\
 \!\!\!\mathrm{s.t. }\qquad\,\,& (\ref{eq:problem-original-power}), (\ref{eq:convexified-fronthaul}), \nonumber 
\end{align}
\end{subequations}
\endgroup
with $\boldsymbol{\tau}^{\text{dl}} \!=\! \{\tau_k^{\text{dl}}\}_{k\in\mathcal{K}_U}$, $\boldsymbol{\omega}^{\text{dl}} \!=\! \{\omega_k^{\text{dl}}\}_{k\in\mathcal{K}_U}$, and $\boldsymbol{\Xi}^{\text{dl}} \!=\! \{\boldsymbol{\Xi}_i^{\text{dl}}\}_{i\in\mathcal{K}_A}$.

Keeping the auxiliary variables $\{\boldsymbol{\tau}^{\text{dl}},\boldsymbol{\omega}^{\text{dl}},\boldsymbol{\Xi}^{\text{dl}}\}$ fixed, the optimization of $\{\mathbf{v}^\text{dl}, \boldsymbol{\Omega}^\text{dl}\}$ reduces to a convex problem solvable with standard optimization solvers.
Conversely, keeping $\{\mathbf{v}^\text{dl}, \boldsymbol{\Omega}^\text{dl}\}$ fixed, the optimal auxiliary variables $\{\boldsymbol{\tau}^{\text{dl}},\boldsymbol{\omega}^{\text{dl}},\boldsymbol{\Xi}^{\text{dl}}\}$ are obtained in closed form as given in (\ref{eq:opt-auxiliary-tau-dl}), (\ref{eq:opt-auxiliary-omega-dl}) and (\ref{eq:opt-auxiliary-Xi-dl}).
Thus, by alternately optimizing $\{\mathbf{v}^\text{dl}, \boldsymbol{\Omega}^\text{dl}\}$ and $\{\boldsymbol{\tau}^{\text{dl}},\boldsymbol{\omega}^{\text{dl}},\boldsymbol{\Xi}^{\text{dl}}\}$, we can achieve a sequence of non-decreasing objective values.
The algorithmic details are provided in Sec. \ref{sub:overall-algorithm-downlink}.


\subsection{Optimization of Wave-Domain Post-Processing} \label{sub:opt-wave-downlink}


In this subsection, we optimize the wave-domain post-processing variables $\boldsymbol{\theta}^\text{dl}$ while keeping $\{\mathbf{v}^\text{dl}, \boldsymbol{\Omega}^\text{dl}\}$ fixed. It is noted that the element-wise range constraint (\ref{eq:problem-original-wave-range}) on the phase variables $\boldsymbol{\theta}^\text{dl}$ can be disregarded during optimization, as any phase value $\theta_{i,l,m}^{\text{dl}}$ violating (\ref{eq:problem-original-wave-range}) can be projected back onto the feasible range by adding an integer multiple of $2\pi$ without affecting the objective function.
Since $\boldsymbol{\theta}^\text{dl}$ is not subject to the fronthaul constraint (\ref{eq:problem-original-FH}) or power constraint (\ref{eq:problem-original-power}), we employ a gradient ascent (GA) algorithm (see, e.g., \cite{Ye:TSP03, Lee:TWC10}).
In this approach, $\boldsymbol{\theta}^\text{dl}$ is iteratively updated in the direction of the steepest increase of the objective function, with a step size that decreases gradually over the iterations.

To proceed, we compute the partial derivative of the objective function $f_{\text{obj}}^{\text{dl}} = \sum_{k\in\mathcal{K}_U} \alpha_k^\text{dl} R_k^{\text{dl}}$ with respect to each phase element $\theta_{i,l,m}^\text{dl}$ in the following proposition.
\newtheorem{Proposition}{Proposition}

\begin{proposition}\label{prop:PD}

The partial derivative of $f_{\textnormal{obj}}^{\textnormal{dl}}$ with respect to $\theta_{i,l,m}^\textnormal{dl}$ is given by
\begingroup
\allowdisplaybreaks
\begin{align}
    \frac{\partial f_{\textnormal{obj}}^{\textnormal{dl}}}{\partial \theta_{i,l,m}^\textnormal{dl}} =& \frac{2}{\ln 2} \! \sum\nolimits_{k\in\mathcal{K}_U} \!\!\alpha_k^\textnormal{dl} \delta_k^\textnormal{dl} \bigg( \eta_{k,k,i,l,m}^\textnormal{dl}-\gamma_k^\textnormal{dl} \nonumber \\
    &  \,\,\,\,\,\,\times\left( \sum\nolimits_{k^{\prime}\in\mathcal{K}_U\setminus\{k\}} \!\!\eta_{k,k^{\prime},i,l,m}^\textnormal{dl} + \zeta_{k,i,l,m}^\textnormal{dl} \!\right) \!\bigg) \!, \label{eq:PD}
\end{align}
\endgroup
where $\delta_k^\textnormal{dl}$, $\eta_{k,k^{\prime},i,l,m}^\textnormal{dl}$, and $\zeta_{k,i,l,m}^\textnormal{dl}$ are defined as
\begingroup
\allowdisplaybreaks
\begin{subequations} \label{eq:PD-delta-eta-zeta}
\begin{align}
    &\!\!\!\!\delta_k^\textnormal{dl} = 1 / \left( \big|{(\tilde{\mathbf{h}}_k^\textnormal{dl})^H  \mathbf{v}_k^\textnormal{dl}} \big|^2 + \textnormal{IF}_k^{\textnormal{dl}}\big(\mathbf{v}^\textnormal{dl}, \boldsymbol{\Omega}^\textnormal{dl}, \boldsymbol{\theta}^\textnormal{dl}\big) \right), \label{eq:PD-delta} \\
    &\!\!\!\!\eta_{k,k^{\prime},i,l,m}^\textnormal{dl}\!=\!\textnormal{Im}\! \left[ e^{-j\theta_{i,l,m}^\textnormal{dl}} (\mathbf{v}_{k^{\prime}, i}^\textnormal{dl})^H \mathbf{J}_{i,l,m}^\textnormal{dl}
    \mathbf{h}_{k,i}^\textnormal{dl} (\tilde{\mathbf{h}}_k^\textnormal{dl})^H   \mathbf{v}_{k^{\prime}}^\textnormal{dl} \right]\!, \! \label{eq:PD-eta} \\
    &\!\!\!\!\zeta_{k,i,l,m}^\textnormal{dl}\!=\!\textnormal{Im}\! \left[ e^{-j\theta_{i,l,m}^\textnormal{dl}} (\tilde{\mathbf{h}}_{k,i}^\textnormal{dl})^H (\mathbf{\Omega}_i^\textnormal{dl})^H \mathbf{J}_{i,l,m}^\textnormal{dl} \mathbf{h}_{k,i}^\textnormal{dl} \right]\!, \! \label{eq:PD-zeta}
\end{align}
\end{subequations}
\endgroup
with $\mathbf{J}_{i,l,m}^\textnormal{dl} \!=\! (\mathbf{T}_i^\textnormal{dl})^{\!H} \mathbf{a}_{i,l,m}^\textnormal{dl} (\mathbf{b}_{i,l,m}^\textnormal{dl})^{\!H}$.
Here, $\mathbf{a}_{i,l,m}^\textnormal{dl}$ and $(\!\mathbf{b}_{i,l,m}^\textnormal{dl}\!)^H$ represent the $n$th column of the matrix $\mathbf{A}_{i,l}^\textnormal{dl}$ and the $n$th row of the matrix $\mathbf{B}_{i,l}^\textnormal{dl}$, respectively, with
\begingroup
\allowdisplaybreaks
\begin{subequations} \label{eq:wave-matrix-A-B-dl}
\begin{align}
    &\!\!\!\mathbf{A}_{i,l}^\textnormal{dl} \triangleq
    \begin{cases}
        \mathbf{W}_{i,l}^\textnormal{dl} \mathbf{\Phi}_{i,l-1}^\textnormal{dl} \cdots \mathbf{\Phi}_{i,2}^\textnormal{dl} \mathbf{W}_{i,2}^\textnormal{dl} \mathbf{\Phi}_{i,1}^\textnormal{dl}, & \!\!\!\textnormal{if} \,\,l \neq 1, \\
        \mathbf{I}_M, & \!\!\!\textnormal{if} \,\,l=1,
    \end{cases}  \label{eq:wave-matrix-A-dl} \\
    &\!\!\!\mathbf{B}_{i,l}^\textnormal{dl} \!\triangleq \!
    \begin{cases}
        \mathbf{\Phi}_{i,L}^\textnormal{dl} \mathbf{W}_{i,L}^\textnormal{dl} \mathbf{\Phi}_{i,L-1}^\textnormal{dl} \cdots \mathbf{\Phi}_{i,l+1}^\textnormal{dl} \mathbf{W}_{i,l+1}^\textnormal{dl}, & \!\!\!\textnormal{if} \,\,l \neq L, \\
        \mathbf{I}_M, & \!\!\!\textnormal{if} \,\,l=L.
    \end{cases}  \label{eq:wave-matrix-B-dl}
\end{align}
\end{subequations}
\endgroup

\begin{IEEEproof}
\textnormal{Please refer to Appendix \ref{app:proof-gradient-computation}.}
\end{IEEEproof}

\end{proposition}

With the derived gradient in (\ref{eq:PD}), the GA algorithm iteratively updates each phase element as
\begin{align}
    \theta_{i,l,m}^\text{dl} \leftarrow \theta_{i,l,m}^\text{dl} + \mu \left(1 / \big\|\tilde{\boldsymbol{\theta}}_{i,l}^\text{dl}\big\|\right) \left(\partial f_{\text{obj}}^{\text{dl}} / \partial \theta_{i,l,m}^\text{dl} \right), \label{eq:phase-update}
\end{align}
where $\tilde{\boldsymbol{\theta}}_{i,l}^\text{dl} =[ \partial f_{\text{obj}}^{\text{dl}} / \partial \theta_{i,l,1}^\text{dl} \cdots \partial f_{\text{obj}}^{\text{dl}}/\partial \theta_{i,l,M}^\text{dl} ]^T$ stacks the partial derivatives for all phase elements in the $l$th layer of AP $i$.
To prevent gradient explosion or vanishing, the step size $\mu$ is adjusted iteratively as $\mu \leftarrow \beta \mu$ with a decay rate $\beta \in(0,1)$.
The GA algorithm is described in detail in Sec. \ref{sub:overall-algorithm-downlink}.


\subsection{Overall AO Algorithm, Complexity, and Convergence} \label{sub:overall-algorithm-downlink}

\subsubsection{Overall AO Algorithm}

The proposed AO algorithm jointly optimizes the digital processing variables $\{\mathbf{v}^\text{dl},\boldsymbol{\Omega}^\text{dl}\}$ and the wave-domain post-processing variables $\boldsymbol{\theta}^\text{dl}$ through alternating optimization.
Leveraging the optimization methods detailed in the preceding subsections, the complete procedure is summarized in Algorithm 2.
Specifically, the optimization of digital and wave-domain post-processing is performed in Steps 4--7 and 9--17, respectively.

\begin{algorithm}
\caption{Proposed AO algorithm for joint optimization of $\{\mathbf{v}^\text{dl},\boldsymbol{\Omega}^\text{dl}\}$ and $\boldsymbol{\theta}^\text{dl}$ for downlink data transmission}\label{algorithm-1}
\begin{algorithmic}[1]
\State \textbf{initialize:}
\State Set $\{\mathbf{v}^\text{dl}, \boldsymbol{\Omega}^\text{dl}\}$ so that the constraints (\ref{eq:problem-original-FH}) and (\ref{eq:problem-original-power}) are satisfied, and initialize the phase variables $\boldsymbol{\theta}^\text{dl}$ within $[ 0, 2\pi)$ and the outer iteration count $n^{\text{out}}\leftarrow 1$.
    \Repeat
        \Repeat
            \State Update $\{\boldsymbol{\tau}^\text{dl},\boldsymbol{\omega}^\text{dl},\boldsymbol{\Xi}^\text{dl}\}$ with (\ref{eq:opt-auxiliary-tau-dl}), (\ref{eq:opt-auxiliary-omega-dl}) and (\ref{eq:opt-auxiliary-Xi-dl}).
            \State Update $\{\mathbf{v}^\text{dl}, \boldsymbol{\Omega}^\text{dl}\}$ as a solution of the problem (\ref{eq:problem-digital-restated})
            \Statex \quad\quad\quad for fixed $\{\boldsymbol{\tau}^\text{dl},\boldsymbol{\omega}^\text{dl},\boldsymbol{\Xi}^\text{dl}\}$.
        \Until {Converged or $n^{\text{digital}} \geq n_{\max}^{\text{digital}}$
        \Statex \qquad\quad\,\, (Otherwise, set $n^{\text{digital}} \leftarrow n^{\text{digital}}+1$)}
        \State Set $\mu \leftarrow \mu_0$.
        \Repeat
            \State \textbf{for} $(i,m,l)\in\mathcal{K}_A \times \mathcal{M} \times \mathcal{L}$ \textbf{do} \textbf{(in parallel)}
                \State \quad\,\, Compute $\partial f_{\text{obj}}^{\text{dl}} / \partial \theta_{i,l,m}^\text{dl}$ with (\ref{eq:PD}).
                \State \quad\,\, Update $\theta_{i,l,m}^\text{dl}$ with (\ref{eq:phase-update}).
            \State \textbf{end}
            \State Update $\mu \leftarrow \beta \mu$.
        \Until {Converged or $n^{\text{wave}} \geq n_{\max}^{\text{wave}}$ \Statex \qquad\quad\,\, (Otherwise, set $n^{\text{wave}} \leftarrow n^{\text{wave}}+1$)}
    \Until {Converged or $n^{\text{out}} \geq n_{\max}^{\text{out}}$ \Statex \qquad (Otherwise, set $n^{\text{out}} \leftarrow n^{\text{out}}+1$)}
\end{algorithmic}
\end{algorithm}


\subsubsection{Complexity}

The total complexity $C_{\text{total}}^{\text{dl}}$ of Algorithm 2 is given by $C_{\text{total}}^{\text{dl}} = I_{\text{out}}^{\text{dl}} ( C_{\text{digital}}^{\text{dl}} + C_{\text{wave}}^{\text{dl}} )$, where
$C_{\text{digital}}^{\text{dl}}$ and $C_{\text{wave}}^{\text{dl}}$ stand for the complexities of digital and wave-domain post-processing optimization steps, respectively, and $I_{\text{out}}^{\text{dl}}$ is the number of outer iterations required for convergence.
The complexity $C_{\text{digital}}^{\text{dl}}$ associated with the digital-domain optimization is determined by the product of the number of inner iterations and the per-iteration complexity.
The per-iteration complexity is dominated by that of solving the convex problem (\ref{eq:problem-digital-restated}) for fixed $\{\boldsymbol{\tau}^\text{dl},\boldsymbol{\omega}^\text{dl},\boldsymbol{\Xi}^\text{dl}\}$, which is upper bounded by
$\mathcal{O}( n_V^{\text{dl}} ((n_V^{\text{dl}})^3 + n_O^{\text{dl}}) )$ \cite[p. 4]{BTal:LN19},
with $n_V^{\text{dl}} = \mathcal{O}(K_A K_U N + K_A N^2)$ and $n_O^{\text{dl}} = \mathcal{O}(K_A N^2 ( K_A K_U^2 + N ))$.

The complexity $C_{\text{wave}}^{\text{dl}}$ for optimizing the wave-domain post-processing $\boldsymbol{\theta}^\text{dl}$ is given by the number of inner iterations required  for the convergence of the GA algorithm, multiplied by the per-iteration complexity which scales as $\mathcal{O}(K_A^3 K_U^2 L M^3)$.

\subsubsection{Convergence}

The convergence of the FP approach used in the subalgorithm for optimizing the digital processing $\{\mathbf{v}^\text{dl},\boldsymbol{\Omega}^\text{dl}\}$ (Steps 4--7) is established in \cite{Shen:TN19}.
The other subalgorithm, which optimizes the wave-domain post-processing $\boldsymbol{\theta}^\text{dl}$ (Steps 9--17), adopts a GA method, whose convergence is guaranteed under a proper choice of the step size, as shown in \cite{Bazaraa:06}.
Owing to the convergence of both subalgorithms, the overall algorithm, which alternates between these two subalgorithms, converges to a stationary point. The convergence behavior and speed will be illustrated numerically in Sec. \ref{sub:convegence}.

\section{Numerical Results} \label{sec:numerical}

\subsection{Simulation Setup} \label{sub:simulation-setup}

We consider a hexagonal coverage area of radius 100 m \cite[Fig. 1]{Park:CISS14}, where $K_U=6$ UEs are randomly distributed, and $K_A=3$ SIM-equipped APs, which are located at equi-spaced boundary points, and employ sectorized antennas directed toward the center of the coverage area.
Unless stated otherwise, each AP is equipped with $N=2$ RF chains, and the layers of the SIMs consist of $M=16$ meta-atoms arranged in a 4-by-4 uniform planar array.
The channel vector $\mathbf{h}_{k,i}$ is modeled as a correlated Rayleigh fading channel given by $\mathbf{h}_{k,i} \sim\mathcal{CN}(\mathbf{0}, \beta_{k, i} \mathbf{R}_i)$, where $\beta_{k, i} = \beta_0 ( d_{k,i}^{\text{geo}}/d_0 )^{-3}$ represents the pathloss between UE $k$ and AP $i$. Here, $d_{k,i}^{\text{geo}}$ is the distance between UE $k$ and AP $i$. The reference distance and pathloss are set to $d_0 = 30$ m and $\beta_0 = 10$, respectively.
Assuming an isotropic scattering environment with uniformly distributed multipath components, the $(n,n^{\prime})$th element of the spatial covariance matrix $\mathbf{R}_i$ is given by $ \mathbf{R}_i (n,n^\prime) = \text{sinc}\left(2d_{n,n\prime}^{\text{meta}}/\lambda\right)$ \cite{Bjornson:WCL21}, where $\text{sinc}(x) = \text{sin}\,(\pi x)/(\pi x)$, and $d_{n,n\prime}^{\text{meta}}$ denotes the spacing between the meta-atoms.
A carrier frequency of 28 GHz is considered. All APs are equipped with an identical SIM structure, where the thickness of each SIM is $T_{\text{SIM}} = 5\lambda$, and the inter-layer spacing is $d_{\text{Layer}} = T_{\text{SIM}}/L$.
The area of each meta-atom is given by $S = (\lambda / 2)^2$.
Throughout the section, we evaluate the unweighted sum-rate performance.


\subsection{Convergence Behavior} \label{sub:convegence}

\begin{figure}
\centering

\subfloat[Algorithm 1 for the uplink]{\includegraphics[width=0.48\linewidth] {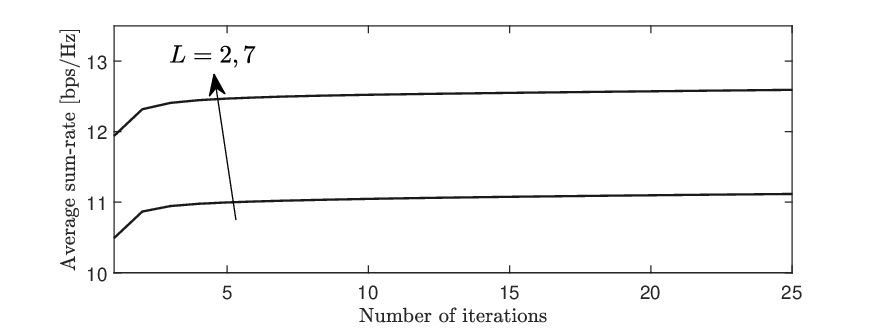}}\hfil
\subfloat[Algorithm 2 for the downlink]{\includegraphics[width=0.48\linewidth] {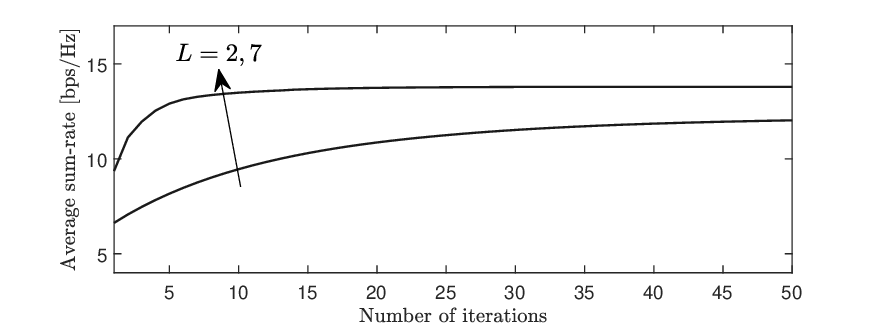}}\hfil
\caption{\small Average sum-rate versus the number of iterations ($K_A=3$, $K_U=6$, $L\in\{2, 7\}$, $M=16$, $P_U/\sigma_{\text{ul}}^2=P_A/\sigma_{\text{dl}}^2=15$ dB, and $C_F=5$ bps/Hz).} \label{fig:graph-vs-iterations}
\vspace{-1mm}
\end{figure}

Fig. \ref{fig:graph-vs-iterations} illustrates the convergence behavior of Algorithms 1 and 2 for uplink and downlink transmissions, respectively, by depicting the average sum-rates versus the number of iterations for $K_A=3$, $K_U=6$, $L\in\{2, 7\}$, $M = 16$, $P_U/\sigma_{\text{ul}}^2 = P_A/\sigma_{\text{dl}}^2=15$ dB, and $C_F=5$ bps/Hz.
The results show that both algorithms exhibit monotonically increasing sum-rates and converge within a few iterations across all simulated scenarios.
Moreover, although the monotonic increase of Algorithm 1 for the uplink is not mathematically guaranteed due to the projection step as discussed in Sec. \ref{sub:overall-algorithm-uplink}, Fig. \ref{fig:graph-vs-iterations}(a) confirms that it exhibits monotonic convergence.



\subsection{Advantages of Hybrid Digital-Wave Scheme in the Uplink} \label{sub:baseline-uplink}

For uplink transmission, we compare the sum-rates of the following baseline and proposed schemes:
\begin{itemize}
    \item \textbf{Fully-digital:}
    Each AP is equipped with $M \gg N$ antennas, not just $N$, each connected to a dedicated RF chain.
    The received signal at each AP $i$'s antennas is thus an $M$-dimensional (not $N$) vector $\mathbf{y}_i^{\text{ul,FD}}$, which is quantized directly without undergoing wave-domaing pre-processing. This results in the quantized signal $\hat{\mathbf{y}}_i^{\text{ul,FD}} \in \mathbb{C}^{M\times 1}$ given by $\hat{\mathbf{y}}_i^{\text{ul,FD}} = \mathbf{y}_i^{\text{ul,FD}} + \mathbf{q}_i^{\text{ul,FD}}$ with the quantization noise vector $\mathbf{q}_i^{\text{ul,FD}} \in \mathbb{C}^{M\times 1}\sim\mathcal{CN}(\mathbf{0}, \boldsymbol{\Omega}_i^{\text{ul,FD}})$.
    In contrast to the proposed hybrid digital-wave scheme, where each AP $i$ only observes the wave-domain pre-processed $N$-dimensional signal $\mathbf{y}_i^{\text{ul}}$, this scheme gives AP $i$ full access to the $M$-dimensional signal $\mathbf{y}_i^{\text{ul,FD}}$ received by its $M$ antennas (i.e., $M$ RF chains). Consequently, this scheme provides a performance upper bound.
    The joint optimization of $\{\boldsymbol{\Omega}_i^{\text{ul,FD}}\}_{i\in\mathcal{K}_A}$ and $\{\mathbf{u}^{\text{ul,FD}}_k\in\mathbb{C}^{M K_A \times 1}\}_{k\in\mathcal{K}_U}$ can be addressed by an AO algorithm similar to Algorithm 1. However, the complexity is significantly higher due to the much larger dimension of the quantization covariance matrices $\boldsymbol{\Omega}_i^{\text{ul,FD}}\in\mathbb{C}^{M\times M}$ compared to $\boldsymbol{\Omega}_i^{\text{ul}}\in \mathbb{C}^{N\times N}$ in the hybrid digital-wave scheme;
    \item \textbf{Hybrid digital-wave (proposed):} The hybrid digital-wave beamforming and fronthaul compression, optimized using Algorithm 1, is applied;
    \item \textbf{Hybrid digital-wave (rand. $\boldsymbol{\theta}^{\textnormal{ul}}$):} The hybrid digital-wave processing is applied, but the SIM phases $\boldsymbol{\theta}^{\text{ul}}$ are randomly fixed. The digital processing $\{\mathbf{p}^{\text{ul}},\boldsymbol{\Omega}^{\text{ul}}, \mathbf{u}^{\text{ul}}\}$ are optimized using Algorithm 1, excluding Steps 4--15;
    \item \textbf{Wave-only:} Beamforming is performed solely through wave beamforming $\boldsymbol{\theta}^{\text{ul}}$, while the digital combiners $\mathbf{u}^{\text{ul}}$ are constrained to
    $\mathbf{u}_k^{\text{ul}} \in \mathbb{R}^{N K_A\times 1}_+$, limiting their role to receive power control.
\end{itemize}


\begin{figure}
\centering\includegraphics[width=0.7\linewidth]{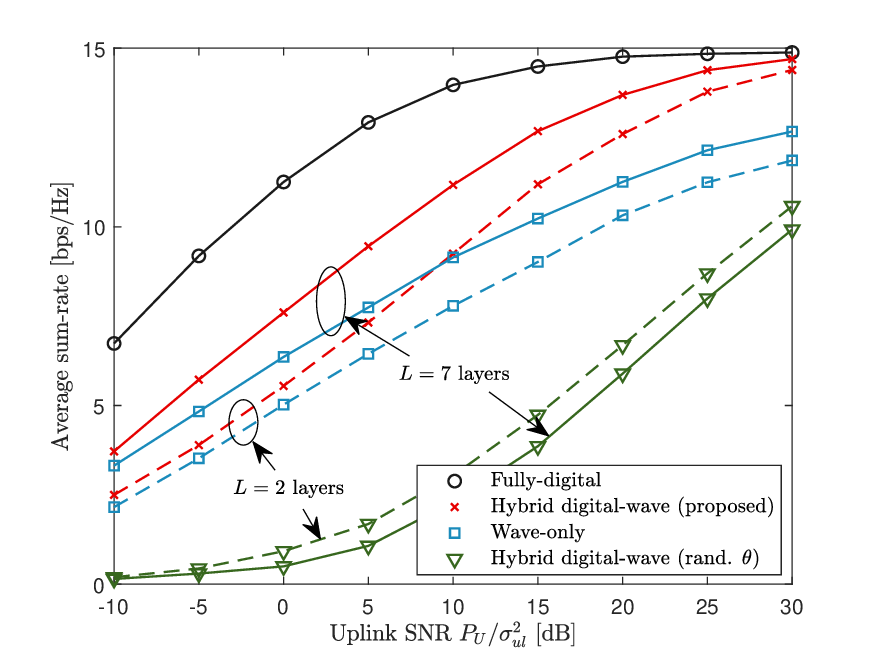}
\vspace{-4mm}
\caption{\small Average sum-rate versus the SNR $P_U/\sigma_{\text{ul}}^2$ for the uplink of SIM-enabled CF-mMIMO systems ($K_A=3$, $K_U=6$, $L\in\{2,7\}$ and $C_F=5$ bps/Hz).} \label{fig:graph-vs-SNR-ul}
\end{figure}

In Fig. \ref{fig:graph-vs-SNR-ul}, we plot the average sum-rate as a function of the transmit SNR level $P_U/\sigma_{\text{ul}}^2$ for $K_A=3$, $K_U=6$, $L\in\{2,7\}$ and $C_F=5$ bps/Hz.
The figure shows that in the low SNR regime, optimizing wave-domain processing has a greater impact than optimizing digital combining vectors due to the higher degrees of control in adjusting the SIM phase shifts $\boldsymbol{\theta}^{\text{ul}}$ compared to adjusting the digital combining vectors $\mathbf{u}^{\text{ul}}$.
Notably, the proposed hybrid digital-wave scheme, using only $N=2$ RF chains, achieves sum-rate performance close to that of the fully-digital scheme with $M=16$ RF chains in the high SNR regime.
Also, the performance gap between the hybrid digital-wave scheme and the fully-digital scheme narrows with an increasing number of layers $L$.
This highlights the potential of SIM to significantly reduce CF-mMIMO system costs while maintaining high sum-rate performance.

\begin{figure}
\centering\includegraphics[width=0.7\linewidth]{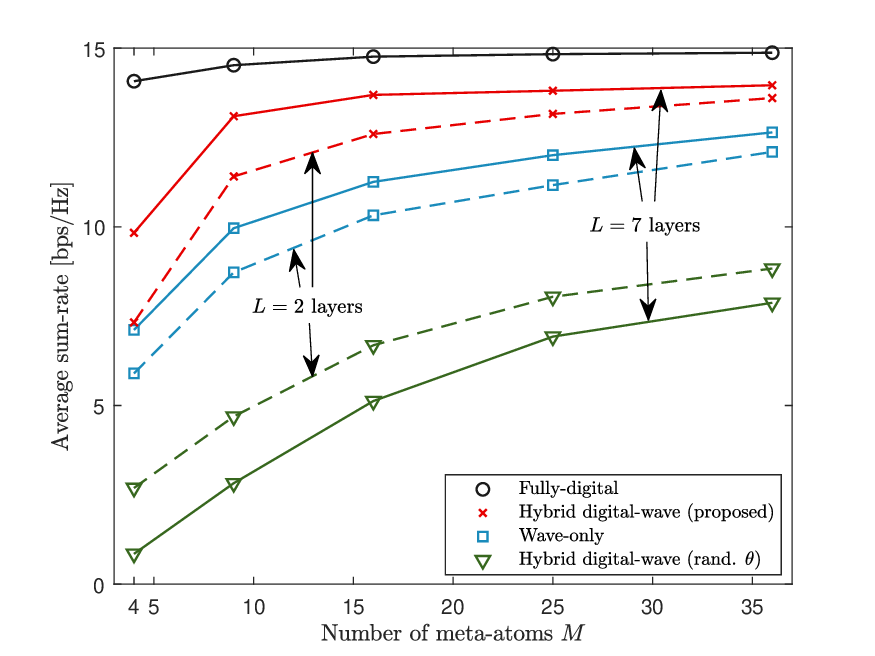}
\vspace{-4mm}
\caption{\small Average sum-rate versus the number of meta-atoms $M$ for the uplink of SIM-enabled CF-mMIMO systems ($K_A=3$, $K_U=6$, $L\in\{2,7\}$, $P_U/\sigma_{\text{ul}}^2=15$ dB, and $C_F=5$ bps/Hz).} \label{fig:graph-vs-metaatoms-ul}
\vspace{-1mm}
\end{figure}

Fig. \ref{fig:graph-vs-metaatoms-ul}  presents the average sum-rate versus the number of meta-atoms $M$ per SIM layer for $K_A=3$, $K_U=6$, $L\in\{2,7\}$, $P_U/\sigma_{\text{ul}}^2=15$ dB, and $C_F=5$ bps/Hz.
The performance gap between the proposed hybrid digital-wave and wave-only schemes remains nearly constant regardless of $M$. However, the performance loss of the hybrid digital-wave scheme with random $\boldsymbol{\theta}^{\text{ul}}$ increases with $M$, since it lacks optimized wave-domain pre-processing.
Additionally, increasing $M$ narrows the sum-rate gap between the proposed hybrid digital-wave and fully-digital schemes, although the gap saturates to a nonzero level. This suggests that a deeper SIM structure is required to fully eliminate the gap as $M$ grows.

\begin{figure}
\centering\includegraphics[width=0.7\linewidth]{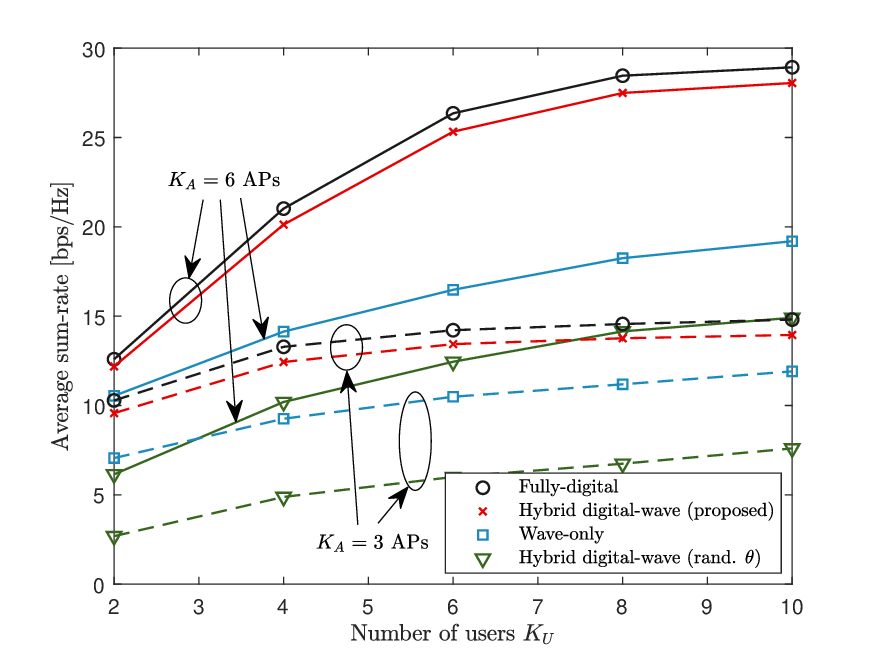}
\vspace{-4mm}
\caption{\small Average sum-rate versus the number of UEs $K_U$ for the uplink of SIM-enabled CF-mMIMO systems ($K_A \in \{3,6\}$, $L=4$, $P_U/\sigma_{\text{ul}}^2=20$ dB, and $C_F=5$ bps/Hz).} \label{fig:graph-vs-UEs-ul}
\vspace{-1mm}
\end{figure}

Fig. \ref{fig:graph-vs-UEs-ul} plots the average sum-rate versus the number of UEs $K_U$ for $K_A \in \{3,6\}$, $L=4$, $P_U/\sigma_{\text{ul}}^2=20$ dB, and $C_F=5$ bps/Hz.
The proposed hybrid digital-wave scheme consistently achieves performance close to that of the fully-digital scheme across the entire range of $K_U$.
For $K_A=3$ APs, the performance gains over the wave-only scheme and the hybrid digital-wave scheme with random $\boldsymbol{\theta}^{\text{ul}}$ slightly diminish as $K_U$ increases, due to the limited degrees of control provided by the SIMs and antennas.
In contrast, with $K_A=6$ APs, the sum-rate gains continue to grow with $K_U$, as the additional degrees of control allow all UEs to be served in an interference-free manner. These results demonstrate that the joint design of digital processing and wave-domain pre-processing becomes increasingly critical in large-scale networks.

In summary, across all simulated scenarios, the wave-only scheme achieves substantial gains over the hybrid digital-wave scheme with random $\boldsymbol{\theta}^{\text{ul}}$, indicating that optimizing the wave-domain pre-processing has a greater impact on overall performance than optimizing the digital processing variables.
Moreover, the proposed hybrid digital-wave scheme notably outperforms the wave-only scheme, attaining sum-rate performance close to the fully-digital bound except in the low-SNR regime.
This suggests that, although the wave-only scheme already improves performance relative to random wave-domain processing scheme, additional gains are realized when wave-domain pre-processing is jointly optimized with digital variables, thereby highlighting the necessity of the proposed joint design.

\subsection{Advantages of Hybrid Digital-Wave Scheme in Downlink} \label{sub:baseline-downlink}

We evaluate and compare the sum-rate performance (i.e., $\alpha_k^{\text{dl}}=1$, $\forall k \in\mathcal{K}_U$) of the following baseline and proposed schemes for the downlink:
\begin{itemize}
    \item \textbf{Fully-digital:} As in the uplink, each AP is equipped with $M \gg N$ antennas, not just $N$, each connected to a dedicated RF chain. The digital-beamformed signal $\mathbf{x}_i^{\text{dl,FD}} = \sum_{k\in\mathcal{K}_U} \mathbf{v}_{k,i}^{\text{dl,FD}}s_k^{\text{dl}}$ and its fronthaul-quantized version $\hat{\mathbf{x}}_i^{\text{dl,FD}} = \mathbf{x}_i^{\text{dl,FD}} + \mathbf{q}_i^{\text{ul,FD}}$ with $\mathbf{q}_i^{\text{dl,FD}}\in\mathbb{C}^{M\times M} \sim \mathcal{CN}(\mathbf{0}, \boldsymbol{\Omega}_i^{\text{dl,FD}})$ are thus $M$-dimensional vectors, and the latter is directly transmitted via AP $i$'s $M$ antennas (i.e., $M$ RF chains) without undergoing wave-domain post-processing.
    Since the digital beamforming vectors $\{\mathbf{v}_{k,i}^{\text{dl,FD}}\in\mathbb{C}^{M\times 1}\}_{k\in\mathcal{K}_U, i\in\mathcal{K}_A}$ are not subject to the wave-domain structural constraints, this fully-digital scheme serves as a performance upper bound. The associated optimization can also be tackled using an AO algorithm. However, its complexity is substantially higher than Algorithm 2 due to the much larger dimensions of the quantization noise covariance matrices;
    \item \textbf{Hybrid digital-wave (proposed):} The hybrid digital-wave beamforming and fronthaul compression, optimized using Algorithm 2, is applied;
    \item \textbf{Hybrid digital-wave (rand. $\boldsymbol{\theta}^{\textnormal{dl}}$):} The hybrid digital-wave processing is applied, but the SIM phases $\boldsymbol{\theta}^{\text{dl}}$ are randomly fixed. The digital processing $\{\mathbf{v}^{\text{dl}},\boldsymbol{\Omega}^{\text{dl}}\}$ are optimized using Algorithm 2, excluding Steps 8--17;
    \item \textbf{Wave-only:} Digital beamforming is limited to power control, constraining each digital beamformer to
    $\mathbf{v}_{k,i}^{\text{dl}}\in\mathbb{R}^{N\times 1}_+$. Consequently, beamforming is exclusively performed through wave beamforming $\boldsymbol{\theta}^{\text{dl}}$.
\end{itemize}

\begin{figure}
\centering\includegraphics[width=0.7\linewidth]{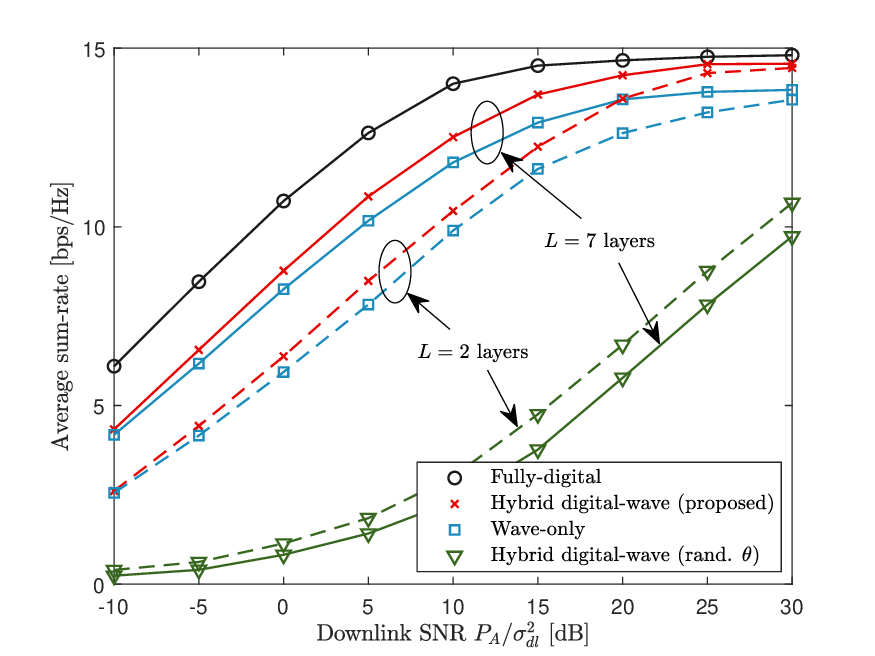}
\vspace{-4mm}
\caption{\small Average sum-rate versus the SNR $P_A/\sigma_{\text{dl}}^2$ for the downlink of SIM-enabled CF-mMIMO systems ($K_A=3$, $K_U=6$, $L\in\{2,7\}$ and $C_F=5$ bps/Hz).} \label{fig:graph-vs-SNR-dl}
\vspace{-1mm}
\end{figure}


In Fig. \ref{fig:graph-vs-SNR-dl}, we depict the average sum-rate while increasing the transmit SNR level $P_A/\sigma_{\text{dl}}^2$ for $K_A=3$, $K_U=6$, $L\in\{2, 7\}$ and $C_F=5$ bps/Hz.
Similar to the uplink results in in Fig. \ref{fig:graph-vs-SNR-ul}, the proposed hybrid digital-wave scheme approaches the fully-digital sum-rate while using only $N=2$ RF chains, in the high SNR regime.
The results also highlight the necessity of joint digital and wave-domain optimization, as the baseline schemes, wave-only and the hybrid-digital scheme with random $\boldsymbol{\theta}^{\text{dl}}$, suffer notable performance degradation.
For the remainder of this subsection, we omit the performance of the hybrid digital-wave scheme with random $\boldsymbol{\theta}^{\text{dl}}$, as it exhibits substantial loss compared to the other schemes.


\begin{figure}
\centering\includegraphics[width=0.7\linewidth]{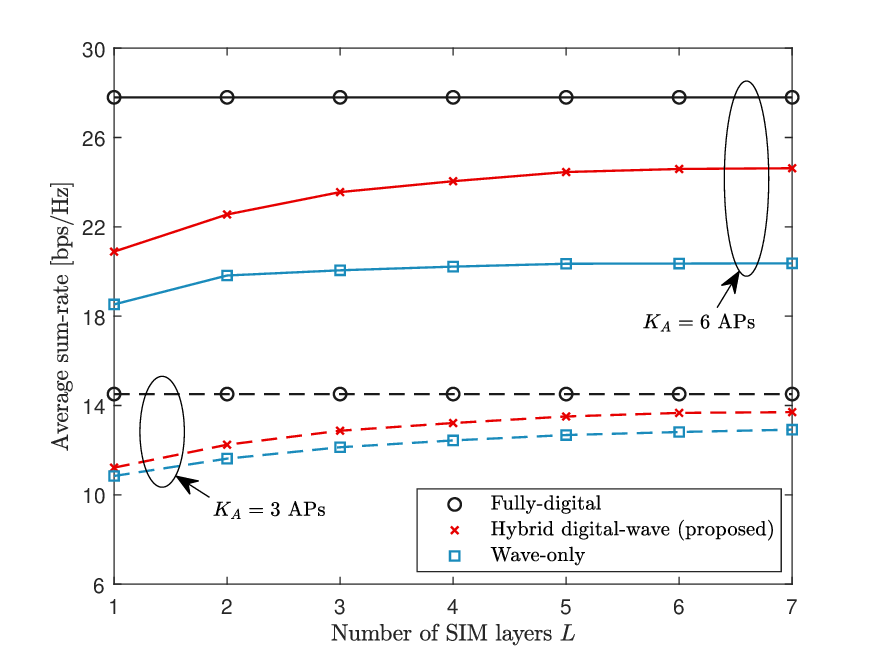}
\vspace{-4mm}
\caption{\small Average sum-rate versus the number of metasurface layers $L$ for the downlink of SIM-enabled CF-mMIMO systems ($K_A\in\{3,6\}$, $K_U=6$, $P_A/\sigma_{\text{dl}}^2=15$ dB and $C_F=5$ bps/Hz).} \label{fig:graph-vs-layers-dl}
\vspace{-1mm}
\end{figure}

Fig. \ref{fig:graph-vs-layers-dl} shows the average sum-rate with respect to the number of metasurface layers $L$ for $K_A\in\{3,6\}$, $K_U=6$, $P_A/\sigma_{\text{dl}}^2=15$ dB and $C_F=5$ bps/Hz.
In the figure, the performance for the hybrid digital-wave scheme with random $\boldsymbol{\theta}$ is excluded, as its observed sum-rates were below 6 bps/Hz.
The sum-rates of both the hybrid digital-wave and wave-only schemes increase with $L$, as the design of wave-domain beamforming benefits from a higher beamforming gain enabled by the larger number of optimization variables.
Furthermore, the performance gap between the proposed hybrid digital-wave scheme and the wave-only scheme increases with $L$ and the number of APs $K_A$.
This highlights the importance of combining wave-domain post-processing enabled by SIM with digital beamforming to achieve performance closer to that of the fully-digital scheme.

\begin{figure}
\centering\includegraphics[width=0.7\linewidth]{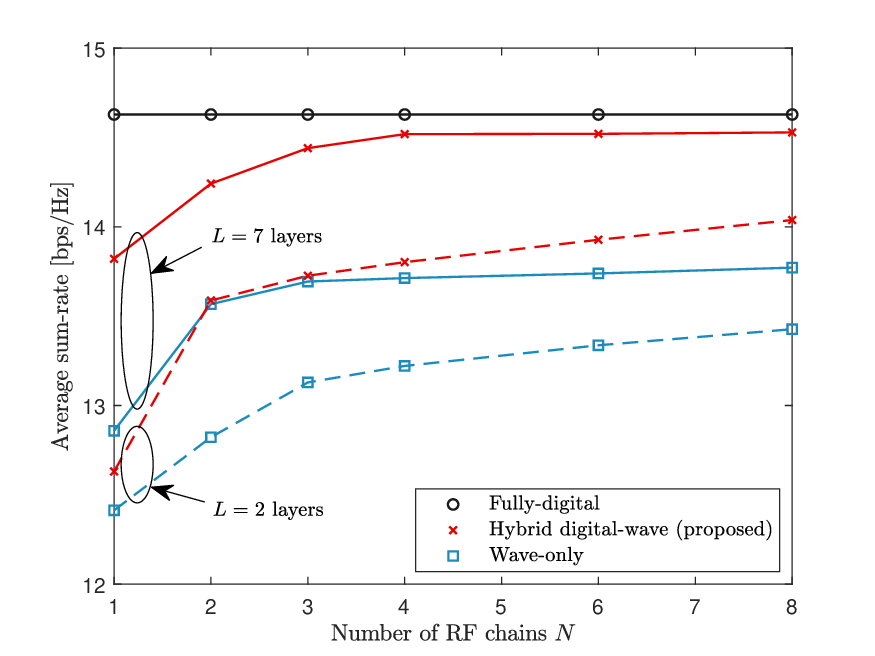}
\vspace{-4mm}
\caption{\small Average sum-rate versus the number of RF chains $N$ for the downlink of SIM-enabled CF-mMIMO systems ($K_A=3$, $K_U=6$, $L\in\{2,7\}$, $P_A/\sigma_{\text{dl}}^2=20$ dB and $C_F=5$ bps/Hz).} \label{fig:graph-vs-RFchains-dl}
\vspace{-1mm}
\end{figure}

In Fig. \ref{fig:graph-vs-RFchains-dl}, we plot the average sum-rate as a function of the number of RF chains $N$ for $K_A=3$, $K_U=6$, $L\in\{2,7\}$, $P_A/\sigma_{\text{dl}}^2=20$ dB and $C_F=5$ bps/Hz.
The wave-only scheme, where digital processing is limited to power control, saturates at a significantly lower sum-rate than the fully-digital scheme, even with $L=7$ SIM layers.
In contrast, the proposed hybrid digital-wave scheme with sufficiently deep SIMs (e.g., $L=7$) achieves a sum-rate close to the fully-digital benchmark using only 4 RF chains, substantially fewer than the $M=16$ RF chains used in the fully-digital scheme.

The overall pattern of the performance gap between the proposed hybrid digital-wave scheme and the baseline schemes is consistent with the uplink results presented in Sec. \ref{sub:baseline-downlink}. In both uplink and downlink, the performance gains over the baseline schemes become more pronounced at higher SNR levels and in larger networks with more APs. Moreover, optimizing the wave-domain post-processing has a greater impact on performance than optimizing the digital processing variables.



\subsection{Complexity Comparison With Fully-Digital Scheme} \label{sub:complexity-comparison}

We have observed that the proposed hybrid digital-wave scheme achieves an average sum-rate close to that of the fully-digital upper bound in most scenarios, provided that sufficiently deep SIMs are employed (e.g., $L=7$). This is a promising result, particularly considering that the hybrid digital-wave scheme requires significantly fewer RF chains ($N\ll M$) compared to the fully-digital counterpart.
Although this reduction greatly lowers the hardware and operating costs, one might assume that jointly optimizing the digital and wave-domain beamforming variables incurs computational complexity comparable to that of the fully-digital design.
In this subsection, we demonstrate that this is not the case by comparing the computational complexity of the two schemes for both the uplink and downlink.

\subsubsection{Asymptotic Complexity}

We recall that for both the uplink and downlink, we proposed AO algorithms that alternately optimize the digital and wave-domain variables.
Compared to the fully-digital scheme, the asymptotic complexity of optimizing the digital processing variables per iteration is reduced from $\mathcal{O}( ( K_A M^2 + K_U )^4 )$ and $\mathcal{O}( K_A^4 M^4( K_U + M )^4 )$ to $\mathcal{O}(( K_A N^2 + K_U )^4$ and $\mathcal{O}( K_A^4 N^4( K_U + N )^4 )$ for the uplink and downlink, respectively.
This reduction is achieved by replacing the parameter $M$ with $N$, where $N\ll M$, leading to substantial decrease in computational complexity.
However, the proposed algorithms include an additional optimization step for the wave-domain variables, whose per-iteration complexity is given by $\mathcal{O}( K_A^3 L M^3 ( K_A ( L^3 M + K_U M ) + K_U^2 ) )$ and $\mathcal{O}(K_A^3K_U^2LM^3)$ for the uplink and downlink, respectively.
Focusing on the dominant scaling with respect to the number of meta-atoms $M$ (i.e., the number of antennas for the fully-digital scheme), the complexity of the wave-domain updates scales as $M^4$ for the uplink and $M^3$ for the downlink.
These scaling behaviors for the additional wave-domain updates are significantly lower than that of the fully-digital scheme, whose complexity scales as $M^8$ for both uplink and downlink.

\subsubsection{Average Algorithm Runtime}

\begin{figure}
\centering

\subfloat[Uplink]{\includegraphics[width=0.48\linewidth] {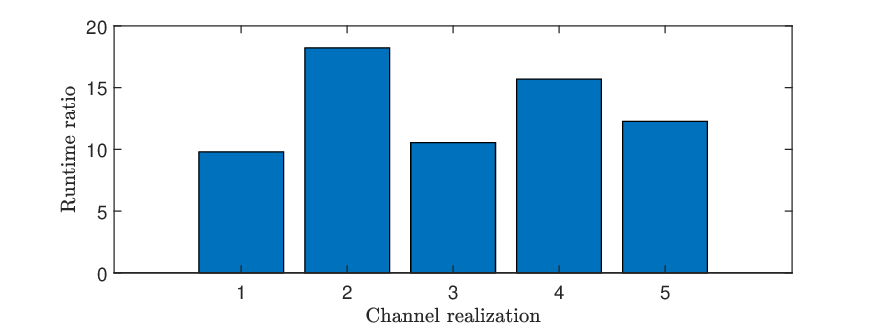}}\hfil
\subfloat[Downlink]{\includegraphics[width=0.48\linewidth] {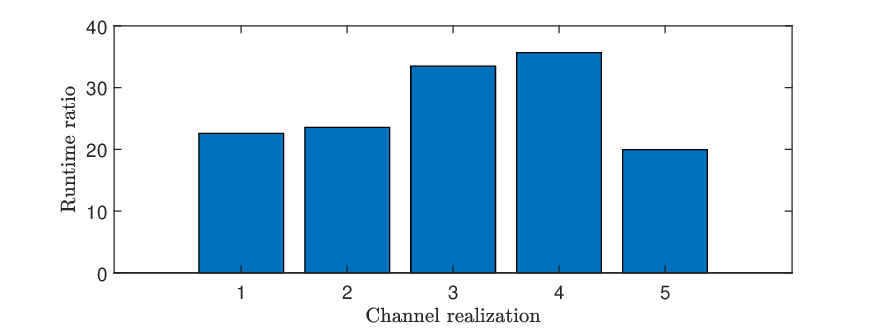}}\hfil
\caption{\small Algorithm runtime ratio for the uplink and downlink transmissions under five independent channel realizations ($K_A=3$, $K_U=6$, $L=7$, $P_U/\sigma_{\text{ul}}^2=P_A/\sigma_{\text{dl}}^2=15$ dB, and $C_F=5$ bps/Hz).} \label{fig:graph-vs-runtime}
\end{figure}

Fig. \ref{fig:graph-vs-runtime} shows the ratio of the algorithm runtime for the fully-digital scheme to that of the proposed hybrid digital-wave schemes in both uplink and downlink, each across 5 independent channel realizations with $K_A=3$, $K_U=6$, $L=7$, and $P_U/\sigma_{\text{ul}}^2=P_A/\sigma_{\text{dl}}^2=15$ dB.
The proposed optimization algorithms reduce the algorithm runtime by more than a factor of 10 and 20, corresponding to over 90 $\%$ and 95 $\%$ savings with respect to time complexity, in the uplink and downlink, respectively.

\subsection{Synergistic Impact of Joint Fronthaul and Wave Beamforming Optimization} \label{sub:impact-opt-fronthaul-compression}

In this subsection, we highlight the significance of optimizing the fronthaul compression strategies $\boldsymbol{\Omega}^{\text{ul}}$ and $\boldsymbol{\Omega}^{\text{dl}}$ particularly in the context of hybrid digital-wave beamforming systems.
To establish a benchmark, we consider an equal-rate compression scheme \cite{Liu:Cambridge17}, where each of the $N$ elements in the uplink and downlink signals, $\bar{\mathbf{y}}^{\text{ul}}_i$ and $\mathbf{x}_i^{\text{dl}}$, is quantized and compressed separately with an equal fronthaul rate allocation of $C_F/N$.
Under this scheme, the quantization noise covariance matrices are constrained to a diagonal form of
$\boldsymbol{\Omega}_i^{X} = \text{diag}\left(\{\nu_{i,n}^X\}_{n\in\mathcal{N}}\right)$, $X\in\{\text{ul}, \text{dl}\}$, where the diagonal elements satisfy the constraints.
\begin{subequations} \label{eq:equal-rate-FH}
\begin{align}
    &\nu_{i,n}^{\text{ul}} \geq \tilde{C}_F \,\mathbf{e}_n^H \left( \sum\nolimits_{k\in\mathcal{K}_U} p_k^\text{ul} \tilde{\mathbf{h}}_{k,i}^\text{ul} (\tilde{\mathbf{h}}_{k,i}^\text{ul})^H + \sigma_\text{ul}^2 \mathbf{I}_N \right)\mathbf{e}_n, \label{eq:equal-rate-FH-uplink} \\
    &\nu_{i,n}^{\text{dl}} \geq \tilde{C}_F \,\mathbf{e}_n^H\left( \sum\nolimits_{k\in\mathcal{K}_U} \mathbf{v}_{k,i}^\text{dl}(\mathbf{v}_{k,i}^\text{dl})^H \right) \mathbf{e}_n, \label{eq:equal-rate-FH-downlink}
\end{align}
\end{subequations}
for all $(i, n)\in\mathcal{K}_A \times \mathcal{N}$.
Here we define $\tilde{C}_F = 1/(2^{C_F/N} - 1)$, while $\mathbf{e}_n \in \mathbb{C}^{N\times 1}$ is a unit vector with its $n$th element equal to 1 and all the other elements set to 0.

\begin{figure}
\centering

\subfloat[Uplink\vspace{-0.5mm}\label{fig:graph-vs-CF-ul}]{\includegraphics[width=0.48\linewidth] {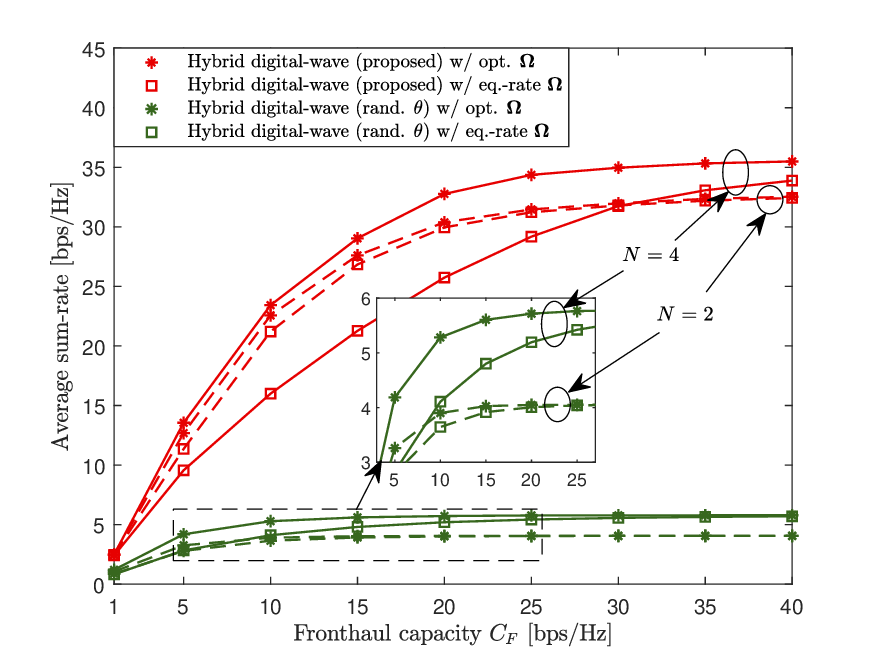}} \hfil
\hspace{0.02\linewidth}
\subfloat[Downlink\label{fig:graph-vs-CF-dl}]{\includegraphics[width=0.48\linewidth] {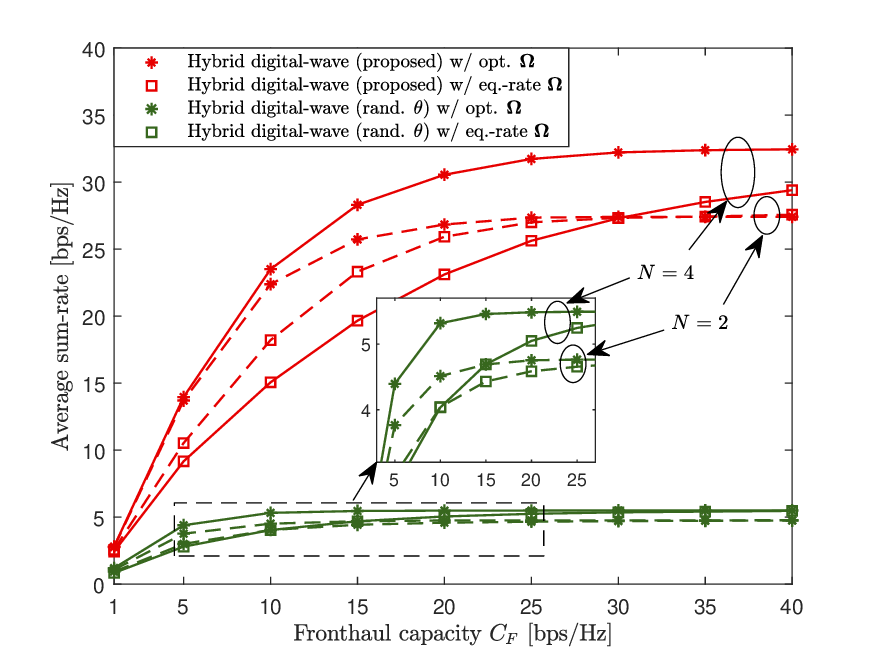}} \hfil

\caption{\small Average sum-rate versus the fronthaul capacity $C_F$ for the uplink and downlink of SIM-enabled CF-mMIMO systems ($K_A=3$, $K_U=6$, $N\in\{2,4\}$, $L=7$, and $P_U/\sigma_{\text{ul}}^2=P_A/\sigma_{\text{dl}}^2=15$ dB).} \label{fig:graph-vs-CF}
\vspace{-1mm}
\end{figure}

Fig. \ref{fig:graph-vs-CF} plots the average sum-rate versus the fronthaul capacity $C_F$ for both uplink and downlink transmissions with $K_A=3$, $K_U=6$, $N\in\{2,4\}$, $L=7$, and $P_U/\sigma_{\text{ul}}^2=P_A/\sigma_{\text{dl}}^2=15$ dB.
The results indicate that optimizing the fronthaul compression strategy, specifically joint compression with adaptive fronthaul rate allocation across $N$ elements, yields greater performance gains when the source signals $\bar{\mathbf{y}}^{\text{ul}}_i$ and $\mathbf{x}_i^{\text{dl}}$ have a larger dimension $N$.
Additionally, these gains are more pronounced when fronthaul compression is jointly optimized with SIM-enabled wave beamforming rather than with randomly fixed SIM phases.
This comparison underscores the critical role of fronthaul compression optimization in hybrid digital-wave beamforming systems, highlighting its greater impact compared to hybrid digital-wave scheme relying on randomly fixed wave-domain processing.

\section{Conclusion} \label{sec:conclusion}

We have proposed a novel hybrid digital and wave-domain beamforming framework for CF-mMIMO systems, which integrates wave-domain beamforming enabled by SIM with conventional digital beamforming. This framework effectively addresses the challenges of high system cost and fronthaul capacity demands, particularly when LAAs are used to improve per-AP coverage.
We formulated the problems of jointly optimizing digital and wave-domain beamforming, along with fronthaul compression, aiming to maximize the weighted sum-rate for uplink and downlink transmissions under finite-capacity fronthaul links.
To solve the non-convex problems, we developed efficient AO-based algorithms, which iteratively optimize digital and wave-domain variables.
Extensive numerical results have demonstrated that the proposed hybrid beamforming schemes significantly outperform conventional schemes that rely on randomly set wave-domain beamformers or restrict digital beamforming to simple power control.
Moreover, the proposed schemes employing sufficiently deep SIMs approach fully-digital performance while requiring substantially fewer RF chains in the high SNR regime.
Our analysis of asymptotic complexity and algorithm runtime confirmed that, compared to the fully-digital schemes, the proposed schemes reduce not only the hardware cost associated with RF chains but also the overall computational complexity. Additionally, the benefits of fronthaul compression optimization are most pronounced when it is jointly optimized with wave-domain beamforming, highlighting the strong synergetic gains of their joint design.


For future work, we plan to extend hybrid digital-wave channel estimators \cite{Nadeem:WCNC24, Yao:WCL24} to CF-mMIMO systems and develop robust hybrid beamforming designs under imperfect CSI \cite{Shen:TSP12}.
Additional research directions include integrating reconfigurable antenna techniques, such as parasitic arrays \cite{Deshpande:arXiv25}, extending uplink-downlink duality results \cite{Liu:TIT21} to SIM-based architectures, optimizing inter-layer transmission matrices using flexible intelligent metasurfaces \cite{An:TWC25-FIM, Mursia:TC25} to further enhance the performance of SIM-aided CF-mMIMO systems, and developing a low-complexity design by extending state-of-the-art efficient algorithms such as, e.g., \cite{Bahingayi:arXiv25, Zhao:TSP23}.

\appendices

\section{Proof of Proposition \ref{prop:matrix-Lagrangian-duality}} \label{app:proof-matrix-Lagrangian-duality}

Consider the function $\log_2\left(1 + |b|^2 / a\right)$ for $a\in\mathbb{R}_+$ and $b\in\mathbb{C}$.
According to the matrix Lagrangian duality result in \cite[Thm. 2]{Shen:TN19}, the following bound holds:
\begin{align}
    &\log_2\left(1 + |b|^2 / a\right) \geq \log_2\left(1 + \tau\right) - \frac{1}{\ln 2} \nonumber \\
    &\,\,\,\,\,\,\,\,\,\,\,\,+ \frac{1}{\ln 2} (1+\tau) \left(2\mathrm{Re}\{b^*\omega\} - |\omega|^2(|b|^2 + a)\right), \label{eq:lower-bound-app-proof-matrix-Lagrangian-duality}
\end{align}
for any $\tau\in\mathbb{R}_+$ and $\omega\in\mathbb{C}$.
The bound in (\ref{eq:lower-bound-app-proof-matrix-Lagrangian-duality}) becomes tight, when $\tau = |b|^2 / a$ and $\omega = b / (|b|^2 + a)$.

By substituting $a\leftarrow \text{IF}_k^{\text{ul}}\big( \mathbf{p}^\text{ul}\!, \!\boldsymbol{\Omega}^\text{ul}\!, \!\boldsymbol{\theta}^{\text{ul}}\!, \!\mathbf{u}^{\text{ul}} \big)$ and $b\leftarrow \sqrt{p_k^\text{ul}} \big( \mathbf{u}_k^{\text{ul}} \big)^H\tilde{\mathbf{h}}_k^\text{ul}$ into (\ref{eq:lower-bound-app-proof-matrix-Lagrangian-duality}), we obtain the lower bound in (\ref{eq:convexified-objective-ul}).

\section{Proof of Proposition \ref{prop:PD}} \label{app:proof-gradient-computation}

The partial derivative of $f_{\text{obj}}^\text{dl}$ with respect to $\theta_{i,l,m}^\text{dl}$ can be written as
\begin{align}
    \frac{\partial f_{\text{obj}}^\text{dl}}{\partial \theta_{i,l,m}^\text{dl}} = \frac{1}{\ln2} \sum_{k \in \mathcal{K}_U} \frac{\alpha_k^\text{dl}}{1+\gamma_k^\text{dl}} \frac{\partial \gamma_k^\text{dl}}{\partial \theta_{i,l,m}^\text{dl}}.  \label{pf:PD1}
\end{align}
Following the standard quotient rule for derivative, $\partial \gamma_k^\text{dl} \!/\! \partial \theta_{i,l,m}^\text{dl}\!$ is given as (\ref{pf:PD2}) shown at the top of this page.
\begin{figure*}[!t]

\begin{align}
    \frac{\partial \gamma_k^\text{dl}}{\partial \theta_{i,l,m}^\text{dl}}
    = &\frac{1}{\text{IF}_k^{\text{dl}}\big(\mathbf{v}^\text{dl}, \boldsymbol{\Omega}^\text{dl}, \boldsymbol{\theta}^\text{dl}\big)}
    \frac{\partial \left| (\tilde{\mathbf{h}}_k^\text{dl})^H \mathbf{v}_k^\text{dl} \right|^2}{\partial \theta_{i,l,m}^\text{dl}} \nonumber \\
    & - \frac{\left| (\tilde{\mathbf{h}}_k^\text{dl})^H \mathbf{v}_k^\text{dl} \right|^2}{\left( \text{IF}_k^{\text{dl}}\big(\mathbf{v}^\text{dl}, \boldsymbol{\Omega}^\text{dl}, \boldsymbol{\theta}^\text{dl}\big) \right)^2}
    \!\left[ \sum_{k^{\prime}\in\mathcal{K}_U\setminus\{k\}} \frac{\partial \left| (\tilde{\mathbf{h}}_k^\text{dl})^H \mathbf{v}_{k^{\prime}}^\text{dl} \right|^2
    }{\partial \theta_{i,l,m}^\text{dl}} + \frac{\partial \left\{
    (\tilde{\mathbf{h}}_k^\text{dl})^H \bar{\boldsymbol{\Omega}}^{\text{dl}} \tilde{\mathbf{h}}_k^\text{dl} \right\}}{\partial \theta_{i,l,m}^\text{dl}} \right] \label{pf:PD2}
\end{align}

\hrulefill
\vspace{-3mm}
\end{figure*}

Substituting (\ref{pf:PD2}) into (\ref{pf:PD1}) leads to
\begin{align}
    &\frac{\partial f_{\text{obj}}^\text{dl}}{\partial \theta_{i,l,m}^\text{dl}} = \frac{1}{\ln2} \sum_{k \in \mathcal{K}_U} \alpha_k^\text{dl} \delta_k^\text{dl} \Bigg( \frac{\partial \left| (\tilde{\mathbf{h}}_k^\text{dl})^H \mathbf{v}_k^\text{dl} \right|^2}{\partial \theta_{i,l,m}^\text{dl}} - \gamma_k^\text{dl} \label{pf:PD3} \\
    & \times \Bigg( \sum_{k^{\prime}\in\mathcal{K}_U\setminus\{k\}} \frac{\partial \left| (\tilde{\mathbf{h}}_k^\text{dl})^H \mathbf{v}_{k^{\prime}}^\text{dl} \right|^2
    }{\partial \theta_{i,l,m}^\text{dl}} + \frac{\partial \left\{
    (\tilde{\mathbf{h}}_k^\text{dl})^H \bar{\boldsymbol{\Omega}}^{\text{dl}} \tilde{\mathbf{h}}_k^\text{dl} \right\}}{\partial \theta_{i,l,m}^\text{dl}} \Bigg) \Bigg) \nonumber,
\end{align}
where $\delta_k^\text{dl}$ is defined in (\ref{eq:PD-delta}).

Noting that the effective channel $\tilde{\mathbf{h}}_{k,i}^\text{dl}$ is an affine function of $e^{j\theta_{i,l,m}^\text{dl}}$,
we can compute the partial derivatives of $| (\tilde{\mathbf{h}}_k^\text{dl})^H \mathbf{v}_{k^{\prime}}^\text{dl} |^2$ and $(\tilde{\mathbf{h}}_k^\text{dl})^H \bar{\boldsymbol{\Omega}}^{\text{dl}} \tilde{\mathbf{h}}_k^\text{dl}$ with respect to $e^{j\theta_{i,l,m}^\text{dl}}$ as
\begingroup
\allowdisplaybreaks
\begin{subequations} \label{pf:PD4-5}
\begin{align}
    &\frac{\partial \left| (\tilde{\mathbf{h}}_k^\text{dl})^H \mathbf{v}_{k^\prime}^\text{dl} \right|^2}{\partial \theta_{i,l,m}^\text{dl}} = \frac{\partial\Big| \sum\limits_{i \in \mathcal{K}_A}(\mathbf{h}_{k,i}^\text{dl})^H \mathbf{G}_i^\text{dl} \mathbf{T}_i^\text{dl} \mathbf{v}_{k^{\prime},i}^\text{dl} \Big|^2}{\partial \theta_{i,l,m}^\text{dl}} \nonumber \\
    & = \frac{\partial \Big| \sum\limits_{m \in \mathcal{M}} e^{j\theta_{i,l,m}^\text{dl}} \sum\limits_{i \in \mathcal{K}_A}(\mathbf{h}_{k,i}^\text{dl})^H \mathbf{b}_{i,l,m}^\text{dl} (\mathbf{a}_{i,l,m}^\text{dl})^H \mathbf{T}_i^\text{dl} \mathbf{v}_{k^{\prime},i}^\text{dl} \Big|^2}{\partial \theta_{i,l,m}^\text{dl}} \nonumber \\
    & = 2 \text{Re} \left[ \left( j e^{j\theta_{i,l,m}^\text{dl}} (\mathbf{h}_{k,i}^\text{dl})^H (\mathbf{J}_{i,l,m}^\text{dl})^H  \mathbf{v}_{k^{\prime},i}^\text{dl} \right) \left( (\tilde{\mathbf{h}}_k^\text{dl})^H \mathbf{v}_{k^\prime}^\text{dl} \right)^H \right] \nonumber \\
    & = 2 \text{Im} \left[ \left( e^{j\theta_{i,l,m}^\text{dl}} (\mathbf{h}_{k,i}^\text{dl})^H (\mathbf{J}_{i,l,m}^\text{dl})^H  \mathbf{v}_{k^{\prime},i}^\text{dl} \right)^H \left( (\tilde{\mathbf{h}}_k^\text{dl})^H \mathbf{v}_{k^\prime}^\text{dl} \right) \right] \nonumber \\
    & = 2 \eta_{k,k^\prime,i,l,m}^\text{dl}, \label{pf:PD4} \\
    &\frac{\partial \left\{
    (\tilde{\mathbf{h}}_k^\text{dl})^H \bar{\boldsymbol{\Omega}}^{\text{dl}} \tilde{\mathbf{h}}_k^\text{dl} \right\}}{\partial \theta_{i,l,m}^\text{dl}} = \frac{\partial \Big\{ \sum\limits_{i \in \mathcal{K}_A}
    (\tilde{\mathbf{h}}_{k,i}^\text{dl})^H \boldsymbol{\Omega}_i^{\text{dl}} \tilde{\mathbf{h}}_{k,i}^\text{dl} \Big\}}{\partial \theta_{i,l,m}^\text{dl}} \nonumber \\
    & = \frac{\partial \Big\{ \sum\limits_{m \in \mathcal{M}} e^{j\theta_{i,l,m}^\text{dl}} \sum\limits_{i \in \mathcal{K}_A}
    (\mathbf{h}_{k,i}^\text{dl})^H \mathbf{b}_{i,l,m}^\text{dl} (\mathbf{a}_{i,l,m}^\text{dl})^H \mathbf{T}_i^\text{dl} \boldsymbol{\Omega}_i^{\text{dl}} \tilde{\mathbf{h}}_{k,i}^\text{dl} \Big\}}{\partial \theta_{i,l,m}^\text{dl}} \nonumber \\
    & = 2 \text{Re} \left[ j e^{j\theta_{i,l,m}^\text{dl}}
    (\mathbf{h}_{k,i}^\text{dl})^H (\mathbf{J}_{i,l,m}^\text{dl})^H \boldsymbol{\Omega}_i^{\text{dl}} \tilde{\mathbf{h}}_{k,i}^\text{dl} \right] \nonumber \\
    & = 2 \text{Im} \left[ e^{-j\theta_{i,l,m}^\text{dl}} (\tilde{\mathbf{h}}_{k,i}^\text{dl})^H (\boldsymbol{\Omega}_i^{\text{dl}})^H
     \mathbf{J}_{i,l,m}^\text{dl} \mathbf{h}_{k,i}^\text{dl} \right] \nonumber \\
    & = 2 \zeta_{k,i,l,m}^{\text{dl}} \label{pf:PD5},
\end{align}
\end{subequations}
\endgroup
where $\eta_{k,k^\prime,i,l,m}^\text{dl}$ and $\zeta_{k,i,l,m}^{\text{dl}}$ are defined in (\ref{eq:PD-eta}) and (\ref{eq:PD-zeta}), respectively.

By substituting (\ref{pf:PD4-5}) into (\ref{pf:PD2}), the proof is completed.

\end{document}